\documentclass[aps,pre,twocolumn,superscriptaddress,floats,floatfix]{revtex4-2}
\usepackage{amssymb,amsmath,amsthm}
\usepackage{graphicx}
\usepackage{epstopdf}
\usepackage{color}
\usepackage{subfigure}
\usepackage{epsfig}
\usepackage{rotating}
\usepackage{mwe}
\usepackage{float}
\usepackage{dcolumn,bm}
\usepackage{verbatim}
\usepackage{hyperref}
\usepackage{pgf,tikz}
 \usetikzlibrary{calc}
\usetikzlibrary{arrows.meta, angles, quotes}
\usepackage{makecell}
\usetikzlibrary{3d}
\usepackage[normalem]{ulem}
\usepackage{xcolor}
\definecolor{Darkred}{RGB}{180, 0, 0}
\definecolor{Darkblue}{RGB}{26,68, 171}
\hypersetup{
  colorlinks = true,     
  urlcolor   = Darkblue, 
  linkcolor  = Darkblue, 
  citecolor  = Darkred   
}

\begin{document}

\begin{abstract}
    We study turbulence in self-gravitating superfluids by performing direct numerical simulations of the 3D Gross-Pitaevskii-Poisson (GPP) equation, which is also a model for dark matter haloes around galaxies. In the absence of self-gravity, the spectrally truncated Gross-Pitaevskii (GP) equation shows the emergence of Kolmogorov's $5/3$ scaling in the incompressible kinetic energy spectrum. Introducing self-gravity, we observe the formation of different structures, from sheet-like to spherically collapsed structures, which introduce a minimum in the kinetic energy spectrum that corresponds to the sizes of these structures. The system shows early convergence towards statistically stationary states, which we show by the onset of thermalisation in the compressible kinetic energy spectrum, where $E_{\rm kin}^c \propto k^2$. We also show that the formation of such large-scale structures suggests that the particles (bosons) move from small to large scales through an inverse cascade, supporting a mechanism for the formation of large-scale structures, such as dark matter haloes, around our galaxy Milky Way.
\end{abstract}

\title{Turbulence and large-scale structures in self-gravitating superfluids}
\author{Sanjay Shukla}
\email{s.shukla@tue.nl}
\affiliation{Department of Applied Physics and Science Education, Eindhoven University of Technology, 5600 MB Eindhoven, The Netherlands}

\maketitle

\section{Introduction}
Turbulence in classical fluids has long been a captivating and challenging problem for both physicists and mathematicians. Interestingly, turbulence has also been observed in inviscid~\cite{Cyril_2005_PhysRevLett}, low-temperature superfluid, and atomic Bose-Einstein condensates~\cite{J_Maurer_1998,annurev_paoletti,Henn_2009_PRL}. However, the study of turbulence in superfluids is still in its infancy. Superfluid dynamics is characterized by the existence of quantum vortices with the quantum of circulation $k=h/m$, where $h$ is the Planck constant and $m$ is the mass of atoms. The motion of these quantum vortices and their turbulence is also fascinating because of the wide range of scales it encompasses: from the microscopic motion of quantum vortices, which occurs at scales of the order of $\sim1$ nm in liquid $^4{\rm He}$, to the grand astrophysical scales found in the interiors of neutron stars ($\sim$10 km). 

Recent advances in scientific facilities and computing resources have enabled the use of Gross-Pitaevskii (GP) modelling of superfluids~\cite{Natalia_2014,Kobayashi_2005_PhysRevLett,Kobayashi_2006_PRL,GK_2012_PhysRevE} to study the tangled state of quantum vortices in the laboratory flows, which we call superfluid turbulence. The GP equation, coupled with the Poisson equation, provides a framework for investigating self-gravitating superfluids, which has been instrumental in studying phenomena like pulsar glitches~\cite{AK_verma_2022,Drummond_2017_b,Shukla_2024_PhysRevD}. The Gross-Pitaevskii-Poisson (GPP) equation also finds applications in modeling ultra-light dark matter around galaxies~\cite{chavanis_2011,Madarassy_2015_PhysRevD,Suarez}, and has recently been extended to examine axion dark matter candidates~\cite{chavanis_2018_PRD,Shukla_2024_PhysRevD.109.063009}. Despite its broad utility, studies of turbulence within the GPP framework remain limited, apart from the development of weak wave turbulence theory~\cite{Jonathan_2020_PhysRevA}. In this paper, we study turbulence in self-gravitating superfluids by performing direct numerical simulations of the three dimensional (3D) GPP equation. 

The GP equation, without accounting for self-gravity, has been extensively applied to investigate the motion of quantum vortices and turbulence in superfluids~\cite{Koplik_1993_PRL,Koplik_1996_PRL,Leadbeater_2001_PRL,Kerr_2011_PRL,Rorai_2013,KOBAYASHI2021107579,shukla2024capturereleasequantumvortices}. These quantum vortices undergo reconnections~\cite{FEYNMAN195517,Leadbeater_2001_PRL,Ogawa_2002_JPS} and transform their energy by emitting sound waves, providing dissipation in the system. The direct numerical simulations (DNSs) of the GP equation have demonstrated the emergence of Kolmogorov's like $5/3$ power-law in the kinetic energy spectrum~\cite{Nore_Marc_1997_POF, Nore_1997_PhysRevLett,Araki_2002_PRL,Polanco_2021NatComm}. The $5/3$ scaling~\cite{Frisch_1995} is a tell-tale signature of high Reynolds number classical turbulence, and for superfluid, it is associated with the dissipative decay of kinetic energy stored in quantum vortices, which have a core size of the order of coherence length $\xi$ [e.g. $\xi \sim 10^{-9}\ m$ in liquid $^4 {\rm He}$]. When self-gravity is incorporated into the GP equation, it introduces a new length scale into the system, Jean's length $\lambda_{\rm J}$, above which gravitational collapse occurs. The introduction of the self-gravity affects the particle number and energy spectra in the system.  Jonathan \textit{et al}~\cite{Jonathan_2020_PhysRevA} shows, using the weak-wave turbulence theory of self-gravitating GP equation, that the particle number spectrum follows an inverse cascade and a direct cascade of the kinetic energy spectrum. The inverse cascade is well known in the two-dimensional Navier-Stokes turbulence, where energy flows from small to large scale, forming coherent structures~\cite{Axel_2012_PhysRevE,Boffetta_2012}. In our case of a three-dimensional self-gravitating GP equation, the inverse cascade is associated with the accumulation of particles at small wavenumbers, which leads to the formation of large-scale structures and helps explain the formation of dark matter haloes around galaxies.

The use of the Poisson equation in the GP system offers a broad framework for studying the non-equilibrium evolution of self-gravitating superfluids and their turbulence. We perform direct numerical simulations (DNSs) of the GPP equation and show that the kinetic energy spectrum deviates from that of the GP as the system collapses because of sufficient gravitational strength. In particular, as the system collapses into a spherical shape, there is a development of minimum in the kinetic energy spectrum at the wavenumber $k_{\rm th}$, which is associated with the diffusive scale, $\lambda_{\rm J}$, or size of the spherical condensate. With an increase in the gravitational strength, the size of the condensate becomes smaller and $k_{\rm th}$ shifts towards higher wavenumbers. This behavior has also been observed in the spectrally truncated DNS of the Euler equation by Cichowlas \textit{et al}~\cite{Cyril_2005_PhysRevLett}, where a minimum in the kinetic energy spectrum appears and provides an effective dissipation for the low wave number ($k < k_{\rm th}$) modes and energy flows from large to small scales. In the case of the self-gravitating Gross-Pitaevskii system, the minimum in the energy spectrum is associated with the formation of large-scale structures.  Consequently, the number of particles at wavenumbers above $k_{\rm th}$ (i.e., small scales) decreases and moves to large scales to form large scale structures.  

The remainder of this paper is organised as follows: In Section~\ref{sec:model}, we present the model and numerical scheme followed by the results in Section~\ref{sec:result}. We discuss our conclusions in Section~\ref{sec:conclusion}.

\section{The Model}
\label{sec:model}
\subsection{The equations and numerical method}
At low temperatures, a three-dimensional system of weakly interacting bosons forms a Bose-Einstein condensate (BEC). When these bosons also interact gravitationally, we call it a self-gravitating BEC, which can be described by a complex macroscopic wavefunction $\psi$. The dynamics of this self-gravitating bosonic system are governed by the Gross-Pitaevskii-Poisson (GPP) equation~\cite{AK_verma_2022}
\begin{eqnarray}
    i\hbar \frac{\partial \psi}{\partial t} &=& -\frac{\hbar^2}{2m}\nabla^2 \psi + g|\psi|^2\psi+m\Phi\psi\,; \nonumber\\
    \nabla^2\Phi &=& 4\pi G (m|\psi|^2-\rho_{\rm bg})\,;
    \label{eq:GPE}
\end{eqnarray}
$m$ is the mass of bosons, $g=4\pi a\hbar^2/m$ is the strength of self-interaction, with $a>0$ the s-wave scattering length, $\Phi$ is the gravitational potential, and $G$ is the Newton's gravitational constant. The subtraction of the mean density $\rho_{\rm bg}$ in Eq.~\eqref{eq:GPE} is often called the Jeans swindle~\cite{KIESSLING_2003}, which can be understood by introducing a Newtonian cosmological constant~\cite{Falco_2013}. 

Eqs.~\eqref{eq:GPE} conserve both the total number of particles and the total energy, given as follows
\begin{eqnarray}
    N &=& \int |\psi|^2 {\rm d}{\bf x}\,;\nonumber\\
    E &=& E_{\rm kin} + E_{\rm int} + E_{G}\,,
    \label{eq:conserved_quantity}
\end{eqnarray}
where $E_{\rm kin}$, $E_{\rm int}$, and $E_G$ represent the kinetic energy, interaction energy, and gravitational energy, respectively, and are defined as:
\begin{eqnarray}
E_{\rm kin}  &=&  \frac{\hbar^2}{2m} \int {\rm d} {\bf x} |\nabla \psi |^2 \,;\nonumber \\
E_{\rm int} &=&  \frac{g}{2} \int {\rm d} {\bf x}|\psi|^4 \,; \nonumber \\
E_G &=& 2\pi G m^2 \int {\rm d} {\bf x} |\psi|^2 \nabla^{-2}|\psi|^2\,.
\label{eq:energy}
\end{eqnarray} 

The hydrodynamic representation of Eq.~\eqref{eq:GPE} is given by the Madelung transformation 
\begin{eqnarray}
    \psi({\bf r},t) = \sqrt{\tfrac{\rho({\bf r},t)}{m}}e^{i\phi({\bf r},t)}\,,
    \label{eq:madelung}
\end{eqnarray}
using which the velocity field is given as ${\bf v}({\bf r},t) = \tfrac{\hbar}{m}\nabla \phi({\bf r},t)$. If we use this velocity field, the kinetic energy can be rewritten as $E_{\rm kin} = \int \frac{1}{2}\rho {\bf v}^2 {\rm d}{\bf x}$. Furthermore, the kinetic energy can be decomposed into compressible $E_{\rm kin}^c$ and incompressible $E_{\rm kin}^i$ parts using the Helmholtz decomposition $\sqrt{\rho}{\bf v} = (\sqrt{\rho}{\bf v})^c+(\sqrt{\rho}{\bf v})^i$, with the condition that $\nabla\times (\sqrt{\rho}{\bf v})^c=0$ and $\nabla \cdot (\sqrt{\rho}{\bf v})^i=0$. We can express the two components of kinetic energy in the Fourier space and define the energy spectra. The incompressible and compressible kinetic energy spectra are
\begin{eqnarray}
    E_{\rm kin}^i(k) &=& \frac{1}{2L^3} \int |\mathcal{F}_{\bf k}(\sqrt{\rho}{\bf v})^i|^2{\rm d}\Omega_{\bf k}\,;\nonumber\\
    E_{\rm kin}^c(k) &=& \frac{1}{2L^3} \int |\mathcal{F}_{\bf k}(\sqrt{\rho}{\bf v})^c|^2{\rm d}\Omega_{\bf k}\,,
    \label{eq:kinetic_spectra}
\end{eqnarray}
where $\mathcal{F}_{\bf k}$ is the Fourier transform and $\Omega_{\bf k}$ is the solid angle in spectral space.

The cubic nonlinear term in Eq.~\eqref{eq:GPE} accounts for the local interaction between the bosons, and when compared to the diffusion term, it defines the healing length $\xi$:
\begin{eqnarray}
    \frac{\hbar^2}{2m\xi^2} \sim gn \to \xi &= &\frac{\hbar}{\sqrt{2mgn}}\,,
    \label{eq:length_scale_xi}
\end{eqnarray}
 where $n=|\psi|^2$ is the ground state particle density. The last term in Eq.~\eqref{eq:GPE}, which includes the gravitational potential, introduces a nonlocal interaction. By comparing this nonlocal interaction term with the diffusion term, we obtain the Jeans length scale, which determines the threshold above which gravitational collapse occurs:
\begin{eqnarray}
    \frac{\hbar^2}{2m\lambda_{\rm J}^2} \sim 4\pi m^2 G n\lambda_{\rm J}^2 \to \lambda_{\rm J} &=& \bigg(\frac{\hbar^2}{8\pi Gm^3n}\bigg)^{1/4}\,,
    \label{eq:length_scale_Jean}
\end{eqnarray}
The length $\xi$ is also related to the core of a quantum vortex in superfluids, and $\lambda_{\rm J}$ is Jean's length scale above which gravitational collapse occurs. We can rewrite the GPP~\eqref{eq:GPE} equation in terms of the two length scales in Eqs.~\eqref{eq:length_scale_xi}-\eqref{eq:length_scale_Jean} as
\begin{eqnarray}
    i \frac{\partial \psi}{\partial t} &=& -\alpha\nabla^2 \psi + \beta|\psi|^2\psi+\frac{\alpha}{\lambda_{\rm J}^4}\Phi\psi\,;\nonumber\\
    \nabla^2\Phi &=& (|\psi|^2-1)\,,
    \label{eq:GPE_rescaled}
\end{eqnarray}
where $\alpha=\tfrac{c_s\xi}{\sqrt{2}}$, and $\beta=\tfrac{c_s}{\sqrt{2\xi}}$ with $c_s$ the speed of sound. Using the length scales $\xi$ and $\lambda_{\rm J}$, we define the dimensionless ratio 
\begin{eqnarray}
    l=\frac{\xi}{\lambda_{\rm J}}\,.
    \label{eq:ratio_l}
\end{eqnarray}
For $l\ll 1$, the gravitational strength is not sufficient enough and Eq.~\eqref{eq:GPE_rescaled} reduces to the Gross-Pitaevskii equation, which is used to describe the superfluid $^4{\rm He}$ in the weak interaction limit~\cite{Proukakis_2008}. For $l\gg 1$, the strength of gravitation is large enough for the system to form a spherical compact object, and Eq.~\eqref{eq:GPE_rescaled} reduces to the Schr\"{o}dinger-Poisson equation, which is used to study ultralight dark matter around galaxies~\cite{Sin_PhysRevD_1994}. For intermediate values of $l$, Eqs.~\eqref{eq:GPE_rescaled} offer a framework to explore dynamics, capturing both the dynamics of tangled quantum vortices and the formation of gravitationally collapsed structures. This model not only describes dark matter halos around galaxies but also aids in studying how quantum vortices influence galaxy rotation curves~\cite{Rubin_araa}.

We perform pseudospectral direct numerical simulations (DNS) of Eqs.~\eqref{eq:GPE_rescaled} for different values of the ratio $l$ in a cubic domain, with side $L = 2\pi$ and $N^3=512^3$ collocation points, and periodic boundary conditions in all three spatial directions. We employ Fourier expansion and the $2/3$-rule for dealiasing, i.e., we truncate the Fourier modes by setting $\hat{\psi} \equiv 0$ for $ |{\bf k}| > k_{max}$~\cite{HOU_dealiasing}. We choose the speed of sound to be $c_s=1$ and the coherence length $\xi=1.44dx$, where $dx=L/N$ is the spatial resolution. 

\subsection{Dispersion relation}
\begin{figure}[!hbt]
    \centering
    \includegraphics[scale=0.38]{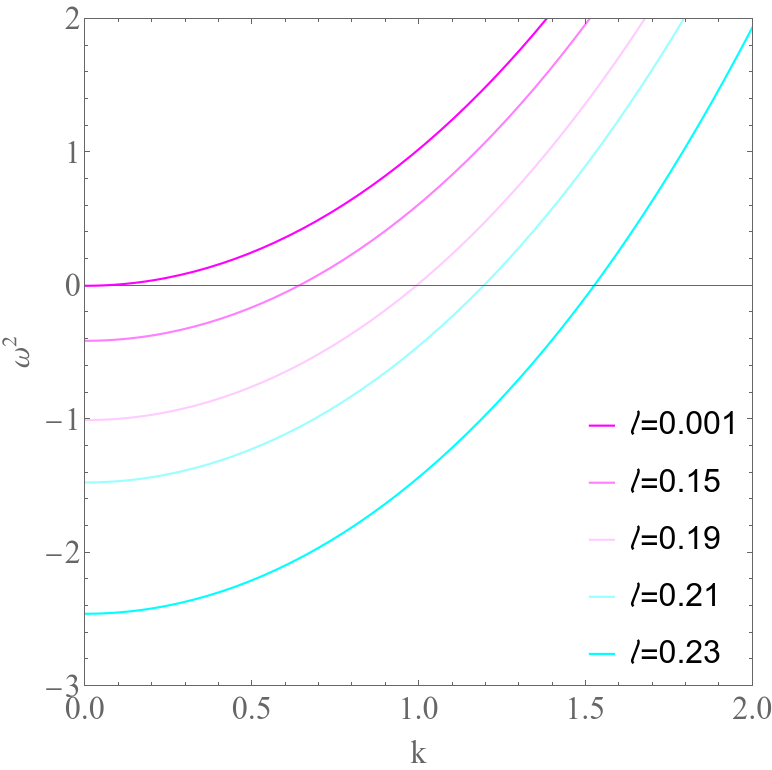}
    \caption{Plots of the dispersion relation in Eq.~\eqref{eq:dispersion} showing $\omega^2$ versus the wavenumber $k$ for different values of the ratio $l=\tfrac{\xi}{\lambda_{\rm J}}$ [Eq.~\eqref{eq:ratio_l}]. The wavenumber where $\omega^2$ passes the $k$-axis gives the Jeans wavenumber $k_{\rm J}$.}
    \label{fig:dispersion}
\end{figure}

We linearise Eq.~\eqref{eq:GPE_rescaled} around the stationary state density $|\psi_0|^2=n_0$ by expressing the wavefunction as $\psi({\bf x},t) = [\psi_0+\delta \psi({\bf x},t)] e^{-i\mu t/\hbar}$, where $\delta \psi$ is the small amplitude perturbation. Assuming a perturbation of the form $\delta\psi = Ae^{i ({\bf k}\cdot {\bf x}-\omega t)}+Be^{-i({\bf k}\cdot {\bf x}-\omega t)}$, we get the dispersion relation, between frequency $\omega$ and wavenumber $k$
\begin{eqnarray}
    \omega = \sqrt{\alpha^2 k^4+2\alpha\beta k^2-\frac{2\alpha^2}{\lambda_{\rm J}^4}}\,,
    \label{eq:dispersion}
\end{eqnarray}

whence we obtain the Jeans wavenumber $k_{\rm J}$, below which the gravitational collapse occurs, by solving for $\omega\equiv0$:
\begin{eqnarray}
k_{\rm J} = \bigg[-\frac{\beta}{\alpha}+\sqrt{\bigg(\frac{\beta}{\alpha}\bigg)^2+\frac{2}{\lambda_{\rm J}^4}}\bigg]^{\frac{1}{2}}\,.
\end{eqnarray}
Fig.~\ref{fig:dispersion} shows the plot of $\omega^2$ vs $k$ from small to large values of the ratio $l=\tfrac{\xi}{\lambda_{\rm J}}$ [Eq.~\eqref{eq:ratio_l}]. For small values of $l$, 
the system does not collapse, and $\omega^2$ is positive for all values of the wavenumber. As we increase the ratio $l$, the system begins to collapse, causing $\omega^2$ to become negative for wavenumbers below the Jeans wavenumber $k_{\rm J}$. Furthermore, with increasing the value of $l$, the wavenumber $k_{\rm J}$ also increases, indicating that the size of the collapsed condensate decreases. 

At the critical value of $l$, where the gravitational force is strong enough to trigger a collapse, a characteristic length scale, $l_{\rm J}\sim k_{\rm J}^{-1}$, emerges within the system. The presence of such a length scale in the system modifies the kinetic energy and particle number spectra of the GPP system. At this value of $l$, when $\omega^2$ starts to become negative, large-scale spherical structures are formed, which occurs due to the accumulation of particles (or bosons) at such scales. At the same time, the number of particles decreases at small scales. This process develops a minimum at wavenumber $k_{\rm th}$ in the kinetic energy spectrum such that for wavenumber $k>k_{\rm th}$, the number of particles decreases. We will now perform DNS of Eqs.~\eqref{eq:GPE_rescaled} to show the change in the kinetic energy spectrum and the emergence of this minimum.

\section{Results}
\label{sec:result}
Through direct numerical simulations, we present a variety of results, such as the formation of different structures and the identification of a minimum in the kinetic energy spectrum as the system collapses into a spherical shape. The formation of a spherical structure occurs through various transitions, such as sheet-like (pancakes) and cylindrical structures. We show the formation of these different structures through the direct numerical simulations (DNSs) of imaginary-time version of Eqs.~\eqref{eq:GPE_rescaled}.

\begin{figure}[!hbt]
    \centering
    \includegraphics[scale=0.38]{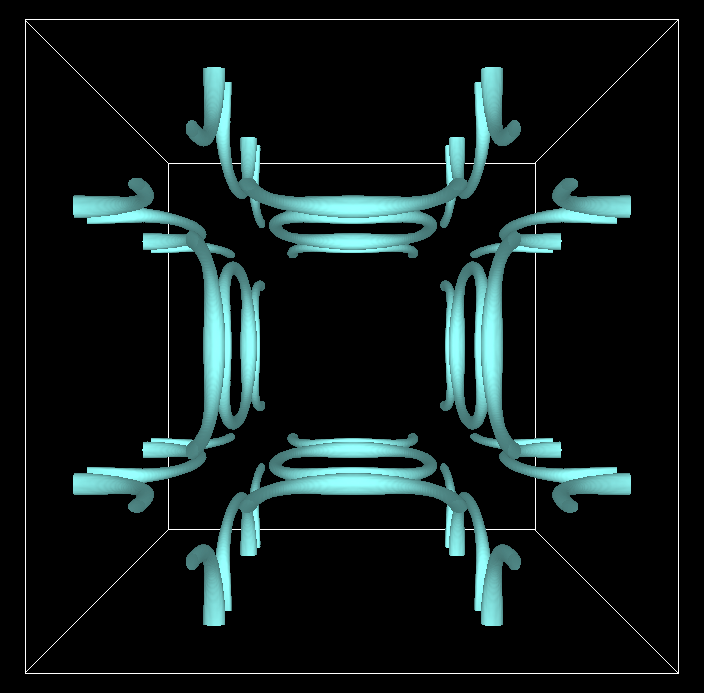}
    \caption{One level contour plots of the density $\rho = m |\psi_{\rm AR}|^2$ describing the Taylor-Green vortex flow from Eq.~\eqref{eq:TG_velocity}.}
    \label{fig:TG_init}
\end{figure}
\begin{figure*}[!hbt]
    \centering
    \includegraphics[scale=0.17]{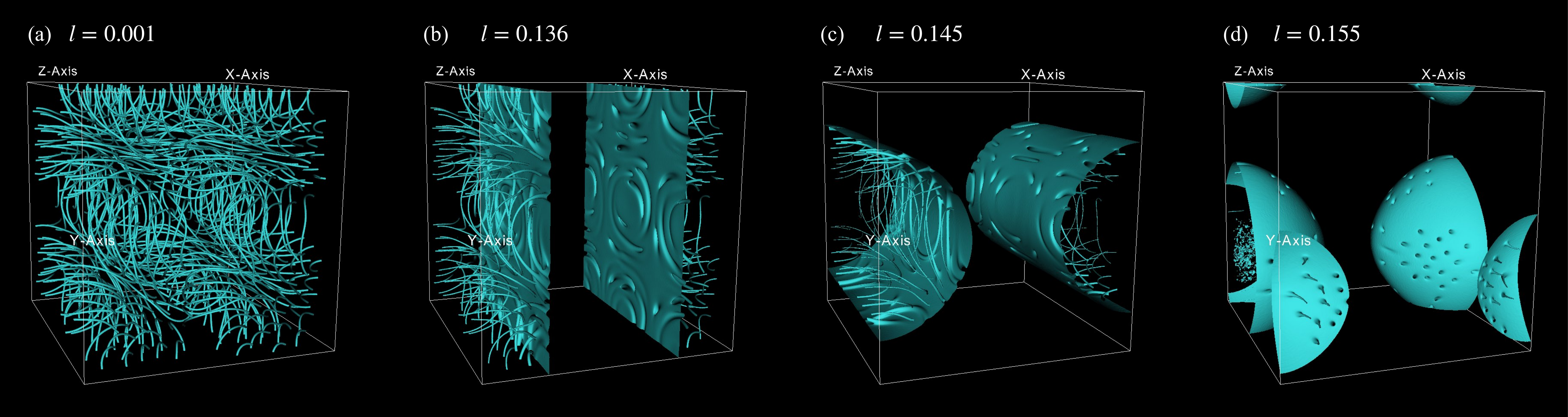}
    \caption{Contour plots of the density $\rho=m|\psi|^2$ in the stationary states after evolving ARGL equation~\eqref{eq:ARGLE} for four values of the ratio $l=\tfrac{\xi}{\lambda_{\rm J}}$: {\bf (a)} $l=0.001$, {\bf  (b)} $l=0.136$, {\bf (c)} $l=0.145$, and {\bf (d)} $l=0.155$.}
    \label{fig:TG_final}
\end{figure*}
\subsection{Numerical simulations and structure formation}
We perform the DNSs of Eqs.~\eqref{eq:GPE_rescaled} using Taylor-Green (TG)
vortex flow~\cite{Taylor_1937}. The TG flow is the solution of the Navier-Stokes equation with the initial velocity field
\begin{eqnarray}
    { v}^{\rm TG}_x&=& \sin(x)\cos(y)\cos(z)\nonumber\,,\\
    { v}^{\rm TG}_y&=&-\cos(x)\sin(y)\cos(z)\nonumber\,,\\
    \text{and} \ { v}^{\rm TG}_z &=& 0\,.
    \label{eq:TG_velocity}
\end{eqnarray}

The generation of the initial condition to solve Eqs.~\eqref{eq:GPE_rescaled}, which corresponds to the flow ${\bf v}^{\rm TG}$, requires multiple steps~\cite{Nore_1997_PhysRevLett}. The first step involves the determination of Clebsch coefficients representing ${\bf v}^{\rm TG}$, which are
\begin{eqnarray}
\lambda(x,y,z)&=&\cos(x)\sqrt{2|\cos(z)|}\nonumber\,,\\
   \text{and} \ \mu(x,y,z)&=& \cos(y)\sqrt{2|\cos(z)|}{\rm sgn}(\cos(z))\,,
\end{eqnarray}
where sgn gives the sign of the argument. We now use these coefficients to construct the wavefunction
\begin{eqnarray}
    \psi_4(\lambda,\mu) = \psi_e(\lambda-1/\sqrt{2},\mu)\psi_e(\lambda,\mu-1/\sqrt{2})\nonumber\\
    \times \psi_e(\lambda+1/\sqrt{2},\mu)\psi_e(\lambda,\mu+1/\sqrt{2})\,,
\end{eqnarray}
where $\psi_e(\lambda,\mu) = \tfrac{(\lambda+i\mu)}{\sqrt{\lambda^2+\mu^2}}\tanh(\tfrac{\sqrt{\lambda^2+\mu^2}}{\sqrt{2}\xi})$. The final wavefunction is given as 
\begin{eqnarray}
    \psi_{\rm AR} = (\psi_4(\lambda,\mu))^{[\gamma/4]}\,,
    \label{eq:ARGLE_initial}
\end{eqnarray}
where $\gamma= \tfrac{2\sqrt{2}}{\pi c_s\xi}$. Fig.~\ref{fig:TG_init} shows the contour plot of density $\rho = m|\psi_{\rm AR}|^2$, describing the TG flow. 

In the second step, we now use $\psi_{\rm AR}$ from Eq.~\eqref{eq:ARGLE_initial} to perform the numerical integration of the advective real Ginzburg-Landau (ARGL) equation, which is the imaginary time ($t\to -i t$) version of Eq.~\eqref{eq:GPE_rescaled} with the advection velocity ${\bf v}^{\rm TG}$
\begin{eqnarray}
     \frac{\partial \psi}{\partial t} &=& \alpha\nabla^2 \psi - \beta|\psi|^2\psi-\frac{\alpha}{\lambda_{\rm J}^4}\Phi\psi-i{\bf v}^{\rm TG}\cdot \nabla \psi-\frac{({\bf v}^{\rm TG})^2}{4\alpha}\psi\,.\nonumber\\
     \label{eq:ARGLE}
\end{eqnarray}

This approach to solve the ARGL in Eq.~\eqref{eq:ARGLE} effectively minimizes the sound waves, allowing the system to quickly reach its ground state and establish a clean initial condition. Using the initial condition $\psi_{\rm AR}$, Fig.~\ref{fig:TG_final} shows the stationary states after evolving Eq.~\eqref{eq:ARGLE} for four values of $l$. For a small value of $l=0.001$, the final state is the uniform random distribution of vortex filaments throughout the box [Fig.~\ref{fig:TG_final}(a)] because of the negligible gravitational strength. As we increase the value of $l$ and reach $l=0.136$, the gravitational strength becomes sufficient and the condensate collapses into a sheet shape as shown in Fig.~\ref{fig:TG_final}(b). With further increase in the value of $l$, the gravitational strength is strong enough and the system collapses into a cylindrical and spherical shape for $l=0.145$ and $l=0.155$, respectively, as shown in Fig.~\ref{fig:TG_final}(c)-(d).

 The collapse of a uniform random distribution of vortices into sheet-like (pancakes), cylindrical, and finally spherical structures is a well-known sequence in gravitational collapse. In the context of cosmological structure formation, this sequence occurs because of the anisotropic nature of gravitational collapse~\cite{Zeldovich}. In our case, involving self-gravitating bosons, the initial vortex distribution is random, and gravitational instability amplifies perturbations differently along each spatial direction. As a result, the collapse first proceeds along the axis with the steepest density gradient, leading to the formation of sheet-like structures [Fig.~\ref{fig:TG_final}(b)]. Subsequently, the instability develops along the other directions, producing cylindrical shapes, and finally completes with a collapse along the third axis, yielding approximately spherical structures. We quantify this directional collapse using the gravitational tidal tensor $T_{ij}$~\cite{PeeblesP}, which describes how a small volume is compressed or stretched along different axes. The components of the tidal tensor are as follows: 
\begin{eqnarray}
    T_{ij}=\frac{\partial^2 \Phi}{\partial x_i  \partial x_j}\,,
    \label{eq:tidal_tensor}
\end{eqnarray}
where $\Phi$ is the gravitational potential given in Eqs.~\eqref{eq:GPE_rescaled}. Using the gravitational potential $\Phi$ obtained from our direct numerical simulations (DNS), we calculate the normalised components of $T_{ij}$ along the three principal axes for four representative values of $l$ given in Fig.~\ref{fig:TG_final} [see Table~\ref{tab:deform}].
\begin{table}[!hbt]
    \centering
    \begin{tabular}{|>{\centering\arraybackslash}p{1.5cm}
                    |>{\centering\arraybackslash}p{1.5cm}
                    |>{\centering\arraybackslash}p{1.5cm}
                    |>{\centering\arraybackslash}p{1.5cm}|}
    \hline
         $l$& $<T_{xx}>$ & $<T_{yy}>$ & $<T_{zz}>$\\
         \hline
         0.001& 0.023& 0.021 &0.019 \\
         0.136& 0.802& 0.017 &0.011 \\
         0.145& 0.823& 0.821 &0.013\\
         0.155& 0.805& 0.787 &0.808 \\
         \hline
    \end{tabular}
    \caption{Component of the tidal tensor $T_{ij}$ in Eq.~\eqref{eq:tidal_tensor} along $x$, $y$, and $z$ directions using the gravitational potential $\Phi$ from our DNSs for uniform random vortices [$l=0.001$], sheet [$l=0.136$], cylinder [$l=0.145$], and sphere [$l=0.155$] in Fig.~\ref{fig:TG_final}.}
    \label{tab:deform}
\end{table}

In Table~\ref{tab:deform}, the notation $<.>$ denotes an average over the entire computational domain. From the table, we observe that for $l=0.136$, $T_{xx} \gg T_{yy},T_{zz}$ indicating that gravitational collapse predominantly occurs along the $x$-direction, resulting in the formation of a sheet-like structure [Fig.~\ref{fig:TG_final}(b)]. A similar analysis for other values of $l$ reveals the progression of collapse: at $l=0.145$ both $T_{xx}$ and $T_{yy}$ are prominent, suggesting a collapse within the $xy$-plane and the emergence of cylindrical structures [Fig.~\ref{fig:TG_final}(c)]. For $l=0.155$, all three components become comparable, indicating that gravitational instability now acts along all directions, leading to the formation of nearly spherical structures [Fig.~\ref{fig:TG_final}(d)].

\begin{figure*}
    \centering
    \includegraphics[scale=0.16]{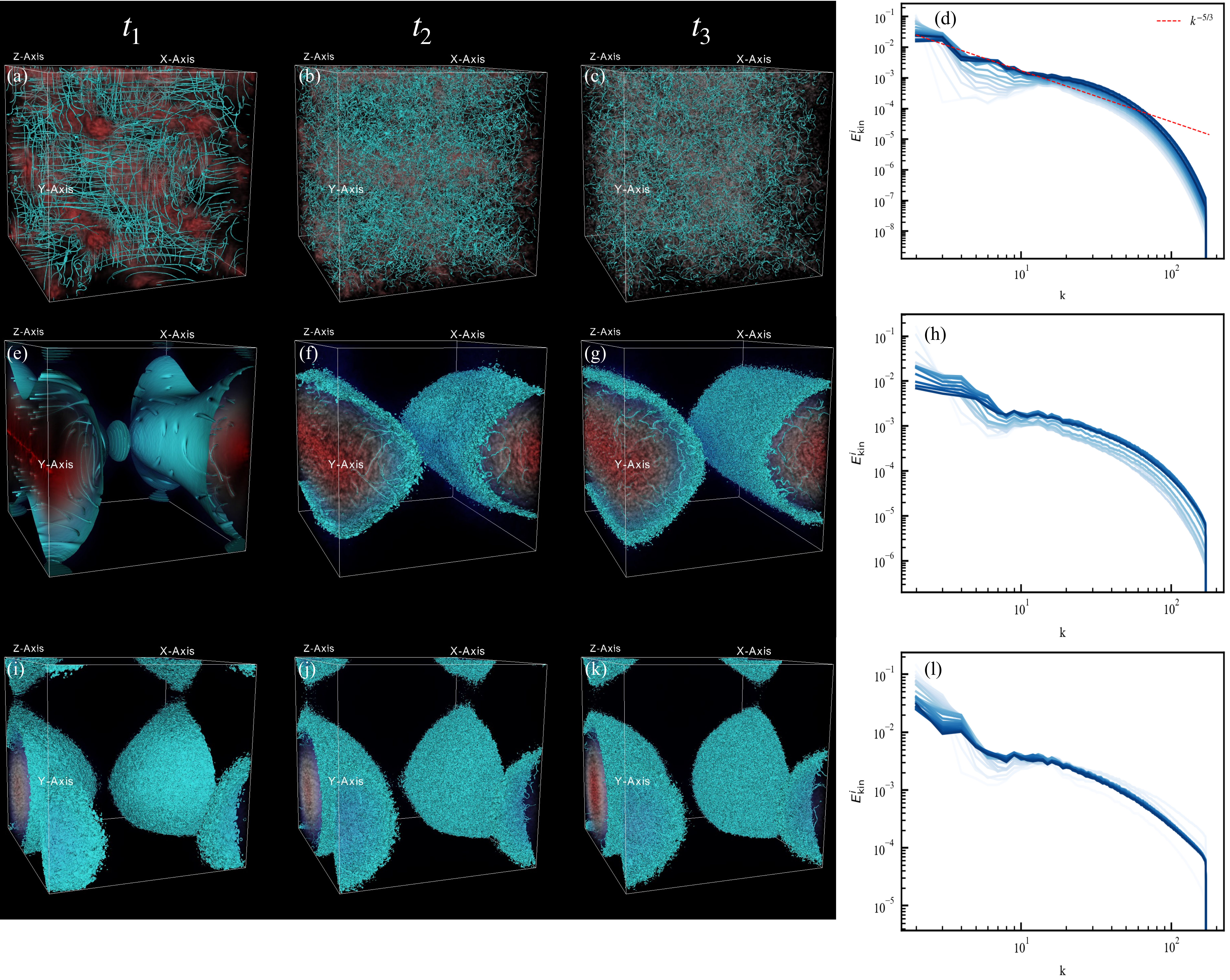}
    \put(-420,355){\large $t_1$}
    \put(-320,355){\large $t_2$}
    \put(-210,355){\large $t_3$}
    \caption{Contour plots of the density $\rho=m|\psi|^2$ from the real-time GPE~\eqref{eq:GPE_rescaled} simulations at three representative times for three different values of the ratio $l=\tfrac{\xi}{\lambda_{\rm J}}$: {\bf (a)}-{\bf (c)} $l=0.001$, {\bf (e)}-{\bf (g)} $l=0.145$, and {\bf (i)}-{\bf (k)} $l=0.155$. Figs.{\bf (d)}, {\bf (h)}, and {\bf (l)} show the incompressible kinetic energy spectra $E_{\rm kin}^i$ [Eq.~\eqref{eq:kinetic_spectra}].}
    \label{fig:rho_GPPE}
\end{figure*}

We now solve the real-time GPP equation~\eqref{eq:GPE_rescaled} using the initial conditions obtained as the stationary state solutions of the ARGL in Eq.~\eqref{eq:ARGLE}. For the initial condition shown in Fig.~\ref{fig:TG_final}(a) at $l=0.001$, the time evolution is given in Figs.~\ref{fig:rho_GPPE}(a)-(c). At this value of $l$, the gravitational strength is not strong enough for a collapse, and we obtain a turbulent tangle of quantum vortices. With the initial condition in Fig.~\ref{fig:TG_final}(c) at $l=0.145$, the system collapses into a cylindrical shape, and the time evolution is shown in Figs.~\ref{fig:rho_GPPE}(e)-(g). For the initial condition in Fig.~\ref{fig:TG_final}(d) at $l=0.155$, the gravitational strength is strong enough for the system to collapse into a spherical shape, and the corresponding time evolution is shown in Figs.~\ref{fig:rho_GPPE}(i)-(k). For real-time dynamics, we omit the sheet-like structure in Fig.~\ref{fig:TG_final}(b) as the length scales are still comparable to the box size and do not significantly affect the energy distribution.

\begin{figure}[!hbt]
    \centering
    \includegraphics[scale=1.2]{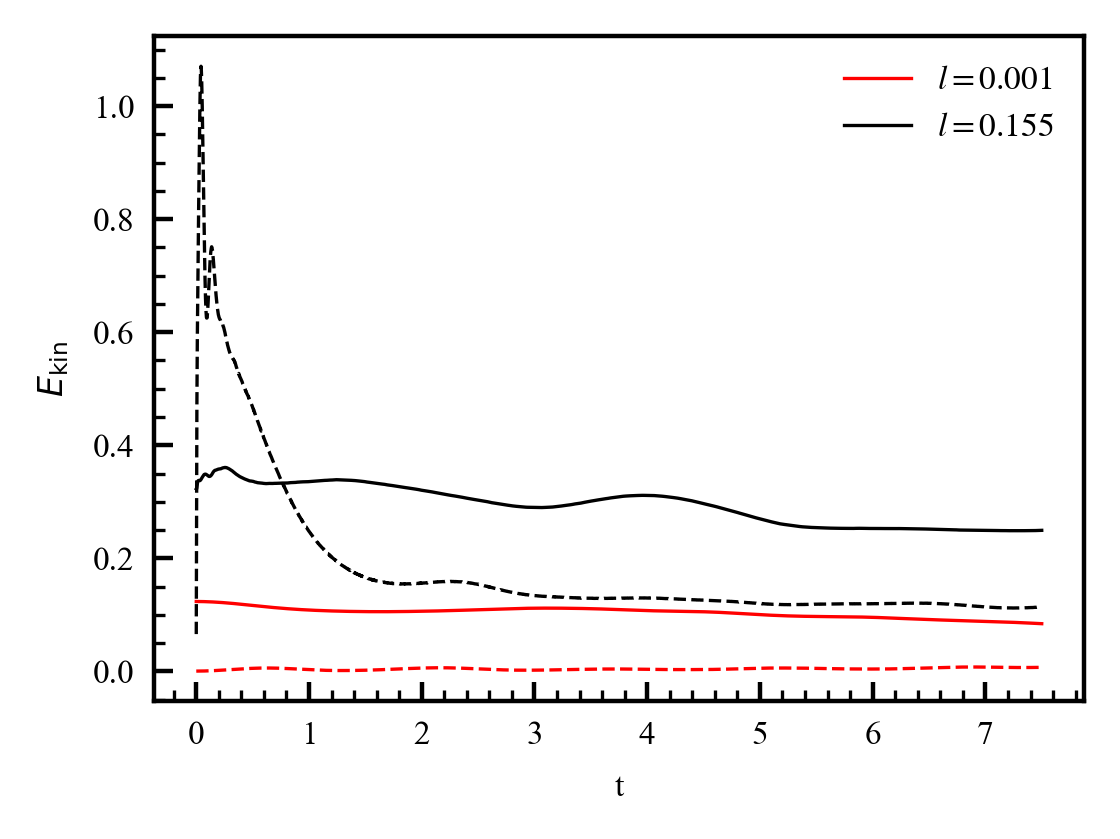}
    \caption{Plots of the incompressible [solid curve] and compressible [dashed curve] kinetic energies versus time [in sec] for the two values of $l=0.001$ and $l=0.155$ corresponding to no collapse and spherical collapse, respectively.}
    \label{fig:inc_com_vs_time}
\end{figure}

After the time evolution of GPP~\eqref{eq:GPE_rescaled}, the system evolves into a tangled, disordered state with different collapsed structures [see Fig.~\ref{fig:rho_GPPE}] for different values of $l$. With increasing $l$, the gravitational strength increases, driving the system toward a spherical shape through different transitions and forming large-scale structures. This introduces a characteristic length scale in the system that modifies the distribution of energy and particles [Eqs.~\eqref{eq:conserved_quantity}] and, hence, the energy spectrum. It is important to note that, as the system collapses to spherical structures, the gravitational potential generates large density gradients from the center to the surface. Because of the large density variations, the gravity contributes primarily to the compressible part of the kinetic energy, which is sensitive to the density variations. We show this by writing the GPP~\eqref{eq:GPE_rescaled} using the Madelung transformation~\eqref{eq:madelung} as follows: 
\begin{eqnarray}
    \frac{\partial \rho}{\partial t} + 2\alpha \nabla \cdot (\rho {\bf v})&=&0\\
    \frac{\partial {\bf v}}{\partial t}+({\bf v}\cdot \nabla){\bf v}&=&-2\alpha \nabla(g\rho + Q) - \frac{2\alpha^2}{\lambda_{\rm J}^4} \nabla\Phi\,,
    \label{eq:gppe_madulung}
\end{eqnarray}
where $Q = -\frac{1}{\sqrt{\rho}}\nabla^2(\sqrt{\rho})$ is the quantum pressure term. If we decompose the velocity field into incompressible and compressible parts ${\bf v} = {\bf v}_i +{\bf v}_c$, we get the following using Eqs.~\eqref{eq:gppe_madulung}
\begin{eqnarray}
    \frac{\partial {\bf v}_{i,c}}{\partial t}&=&-\mathcal{P}_{i,c} \bigg[({\bf v}\cdot \nabla){\bf v}+2\alpha \nabla(g\rho + Q)+\frac{2\alpha^2}{\lambda_{\rm J}^4} \nabla\Phi\bigg]\,,
    \label{eq:inc_com_vel}
\end{eqnarray}
where $\mathcal{P}_i$, $\mathcal{P}_c$ are the projection operators onto the incompressible and compressible parts, respectively. From Eqs.~\eqref{eq:inc_com_vel}, we observe that gravitational potential entered as a gradient $\nabla\Phi$ and we have $\mathcal{P}_i [\nabla\phi] = 0$, $\mathcal{P}_c [\nabla\phi] \neq 0$. This happens because $\nabla\Phi$ is curl-free and the incompressible part of velocity has a non-zero curl. So, during the initial time, the gravitational potential largely contributes to the compressible part of the kinetic energy, while its effect on the incompressible part comes through the non-linear interactions. Fig.~\ref{fig:inc_com_vs_time} shows the time series of incompressible and compressible kinetic energies for two values of $l=0.001$ [no collapse] and $l=0.155$ [spherical collapse] using our DNSs. We observe that for $l=0.001$, the compressible kinetic energy is small with respect to the incompressible one from initial to later times [$t\simeq 8$ sec]. However, for $l=0.155$ where spherical collapse occurs, the compressible energy overshoots the incompressible part during initial evolution and eventually decreases as non-linearity takes place. This confirms our semi-quantitative analysis using Eqs.~\eqref{eq:gppe_madulung}. For these reasons, as the system collapses, the formation of minima in the kinetic energy spectrum is observed dominantly in the compressible part.

To observe the aforementioned effect on energy spectrum, we first calculate the incompressible kinetic energy spectrum $E_{\rm kin}^i(k)$ [Eq.~\eqref{eq:kinetic_spectra}] for three values of $l$ for different time steps. For $l=0.001$ where we have a tangled state of quantum vortices without any collapses, the spectrum $E_{\rm kin}^i(k)$ in Fig.~\ref{fig:rho_GPPE}(d) has the usual behaviour of a 3D quantum turbulence with a Kolmogorov type $-5/3$ power law for wavenumber $k\leq (2\pi)/d$, where $d$ is the average distance between neighbor vortices~\cite{Barenghi_2014_pnas,Nore_1997_PhysRevLett}. The Kolmogorov-like spectrum is a characteristic of the large-scale TG flow. For higher values of $l$, the spectrum of $E_{\rm kin}^i(k)$ does not show Kolmogorov-like power law for $k\leq (2\pi)/d$, as we show in Figs.~\ref{fig:rho_GPPE}(h) and (l). This corresponds to the emergence of gravity-induced large-scale and bound structures, and the spectrum develops a shallow minimum. As we discussed previously, gravitational collapse predominantly affects the compressible part of the kinetic energy as it can not generate vortices; hence, we observe a shallow development of a minimum in the incompressible spectrum.

The incompressible kinetic energy is associated with the motion of quantum vortices. As time progresses, vortices reconnect with each other, and the kinetic energy stored in the motion of quantum vortices is converted into sound waves. Also, for sufficient gravitational strength, the system collapses and generates large density gradients. Fig.~\ref{fig:inc_com_vs_time} shows the plots of incompressible and compressible kinetic energies versus time for $l=0.001$ [no collapse] and $l=0.155$ [spherical collapse], which shows that the collapse primarily affects the compressible part of the kinetic energy at initial times. The interplay of gravitational collapse and kinetic energy develops a minimum in the compressible energy spectrum, whose position depends on the size of large-scale structures.  

\begin{figure*}
    \centering
    \includegraphics[scale=0.15]{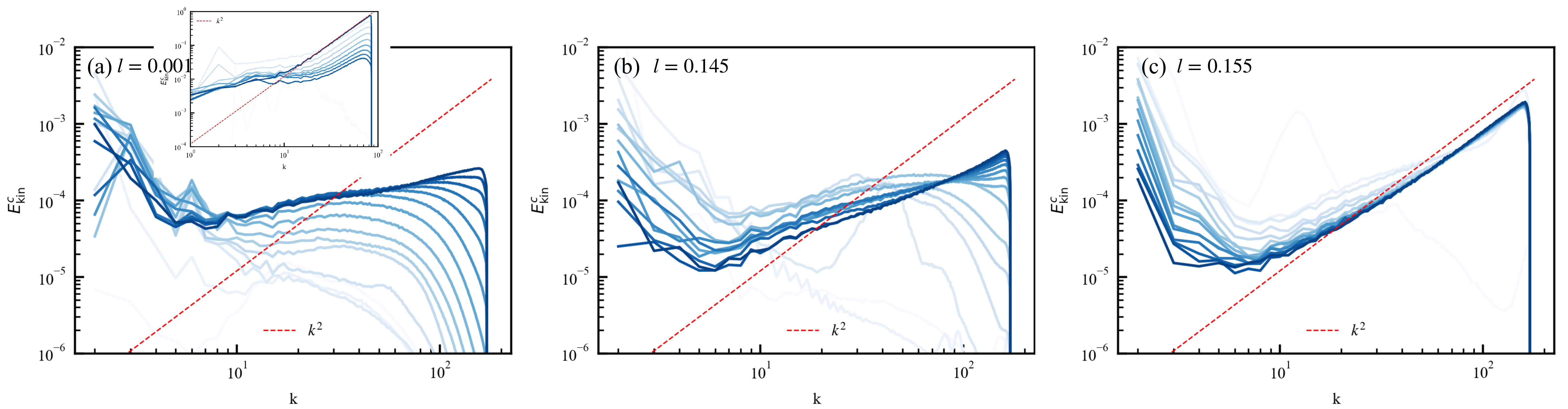}
    \caption{Plots of the compressible kinetic energy spectra $E_{\rm kin}^c$ [Eq.~\eqref{eq:kinetic_spectra}] on log-log scale at different time steps for $l=0.001$ in (a), $l=0.145$ in (b), and $l=0.155$ in (c). The inset in (a) is for resolution of $256^3$, while all other plots are for a resolution of $512^3$.}
    \label{fig:com_spectrum}
\end{figure*}

Fig.~\ref{fig:com_spectrum} shows the compressible kinetic energy spectrum $E_{\rm kin}^c$ at various time steps (up to $t=20$ s for a resolution of $N^3=512^3$) for different values of $l$. For $l=0.001$, Fig.~\ref{fig:com_spectrum}(a) shows that at large wavenumber, the spectrum gradually evolves toward a $k^2$ power-law scaling. We show that the spectrum follows a $k^2$ behavior at large times ($t=90$ s) for a lower resolution of $256^3$ in the inset of Fig.~\ref{fig:com_spectrum}(a). This late time convergence towards $k^2$ indicates thermalisation in the system, which has been discussed in Ref.~\cite{Giorgio_thermalisation} without self-gravity. As we increase the value of $l$, the spectrum exhibits an earlier onset of the $k^2$ scaling, as seen in Figs.~\ref{fig:com_spectrum}(b)-(c). Specifically, Figs.~\ref{fig:com_spectrum}(c) show that the spectrum follows a $k^2$ power law around $t=20$ s. The early-time or accelerated emergence of thermalisation in the system is a consequence of the non-local nature of gravitational interaction. In the absence of gravity, the relaxation time of the Bose-Einstein condensed system follows $\tau_s\propto (g n)^{-1}$~\cite{Kay_PRD_2020}, where $g$ is the coefficient of nonlinear self-interaction, and $n$ is the number density. The use of gravity introduces another timescale into the system based on the Jeans length scale $\lambda_{\rm J}$ [Eq.~\eqref{eq:length_scale_Jean}]. A rough estimate of this time is $\tau_G \simeq \lambda_{\rm J}/c_s$, which in terms of ratio $l$, becomes $\tau_G \simeq \xi /(c_s l) $. A more precise analytical expression of this time scale has been derived in Ref.~\cite{Kay_PRD_2020} following $\tau_G \propto  (G n)^{-1}$, where $G$ is Newton's gravitational constant. In our notations and dimensionless ratio, this time scale becomes 
\begin{eqnarray}
    \tau_G \propto \frac{1}{l^4}\,,
    \label{eq:relaxation}
\end{eqnarray}
where $l$ is the ration given in Eq.~\eqref{eq:ratio_l}. The relaxation time $\tau_G$ because of gravity drastically reduces as we go from $l=0.001$ (no collapse) to $l=0.155$ (spherical collapse). The nonlocal nature of the gravitational interaction couples the fluctuations over the size of the collapsed object and reduces the relaxation time, which appears in the form of accelerated thermalisation in the system. 

\begin{figure}
    \centering
    \includegraphics[scale=1]{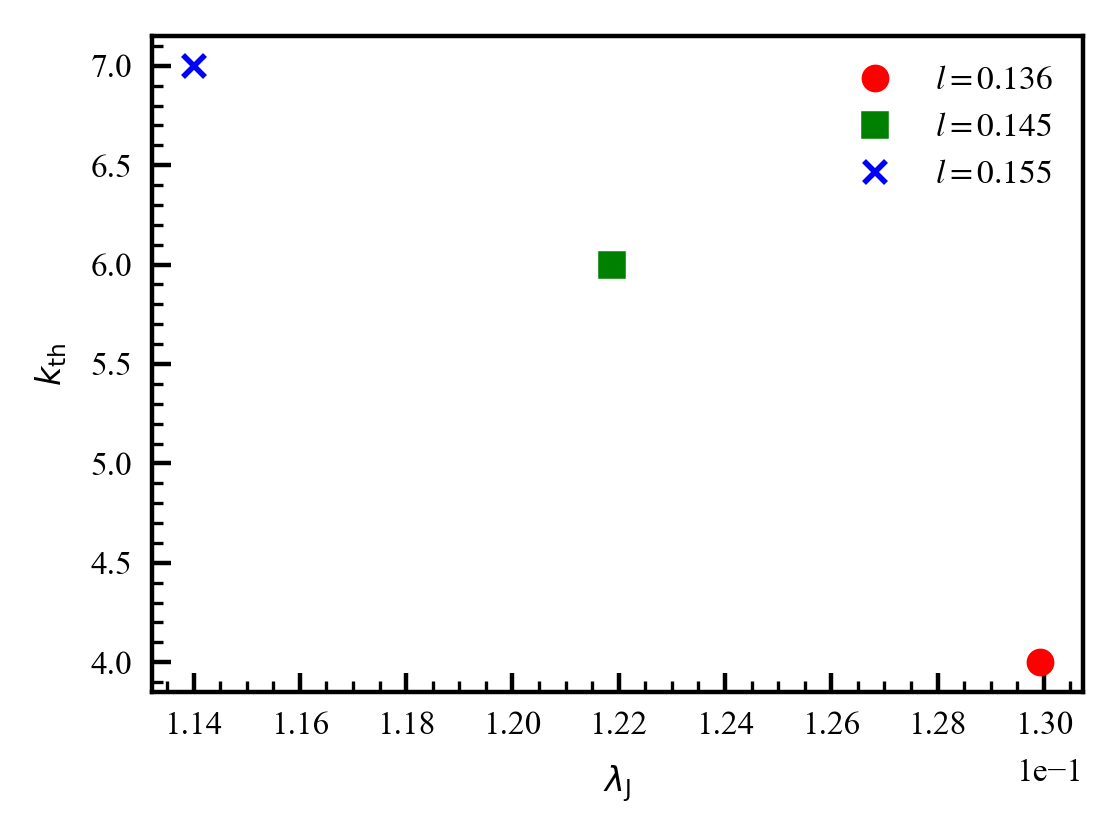}
    \caption{Plot of the threshold wavenumber $k_{th}$, where a minimum appears in the compressible kinetic energy spectrum, versus the diffusive scale $\lambda_{\rm J}$ [Eq.~\eqref{eq:ratio_l}] of the collapsed object.}
    \label{fig:kth_vs_l}
\end{figure}

The goal of the system is to reach a statistical stationary state, which is the solution of the truncated Gross-Pitaevskii-Poisson equation. In this state, all modes are thermalized with the compressible energy spectrum $E_{\rm kin}^c(k)  = ak^2$, where $a$ is a constant. From the compressible kinetic energy spectrum $E_{\rm kin}^c$ shown in Fig.~\ref{fig:com_spectrum}, we observe that the gravitationally collapsed system exhibits accelerated thermalisation, characterised by a $k^2$ power law for wavenumbers $k> k_{th}$. As the system evolves toward a more compact, spherical configuration, the threshold wavenumber $k_{th}$ shifts to higher values. Fig.~\ref{fig:kth_vs_l} shows the plot of $k_{\rm th}$ versus the diffusive scale $\lambda_{\rm J} = \xi/l$ of the collapsed object. For the maximum value of $l$ [spherical object] in our DNS, $\lambda_{\rm J}$ is the smallest and $k_{th}$ is the largest. 

\begin{figure*}
    \centering
    \includegraphics[scale=0.25]{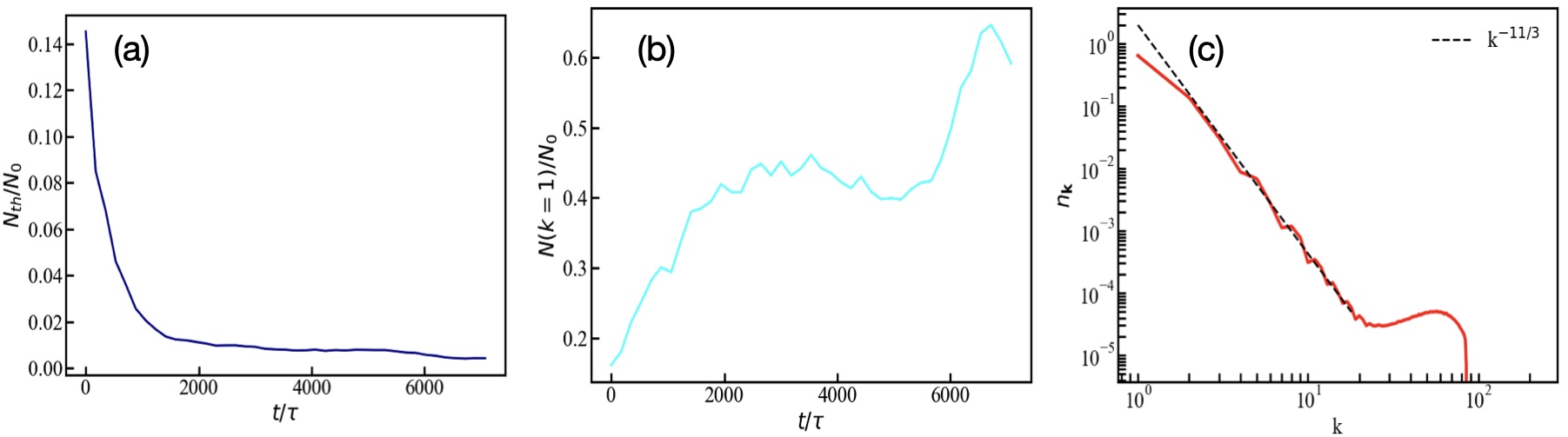}
    \caption{{\bf (a)} Plot of the number of particles, $N_{\rm th}$, versus the scaled time $t/\tau$ for wavenumber $k>k_{\rm th}$. {\bf (b)} Plot of the number of particles versus the scaled time $t/\tau$ in the mode $k=1$ denoted by $N(k=1,t)$. {\bf (c)} Log-Log plot of the partilce number density spectrum $n_{\bf k} = |\psi({\bf k},t)|^2$.}
    \label{fig:part_number}
\end{figure*}

The development of a minimum in the kinetic energy spectrum has consequences on the transfer of the number of particles from small to large scales. The self-gravitating Gross-Pitaevsakii~\eqref{eq:GPE_rescaled} system comes from a conserved Hamiltonian and conserves the number of particles $N = \int |\psi|^2 {\rm d}^3 {\bf x}$~\eqref{eq:conserved_quantity}. We can calculate the decrease in the number of particles above $k_{\rm th}$ with time as
\begin{eqnarray}
    N_{\rm th} = \sum_{k>k_{\rm th}} N(k,t)\,.
\end{eqnarray}
Fig.~\ref{fig:part_number}(a) shows the time evolution of $N_{\rm th}$ that decreases with time. This decrease in the number of particles at large wavenumber is associated with the formation of large-scale gravitationally collapsed structures. Fig.~\ref{fig:part_number}(b) shows the accumulation of the number of particles $N(k=1,t)$ at the largest length scales in our simulations. This suggests that the particles move from small to large scales through an inverse cascade and justifies the mechanism behind the formation of large-scale structures, such as dark matter haloes, around our Milky Way. We also plot the spectrum of particle number density $n_{\bf k} = |\psi({\bf k},t)|^2$ in Fig.~\ref{fig:part_number}(c). The spectrum follows a power law with the exponent $-11/3$ for wavenumbers $k<k_{\rm th}$.

\section{Conclusions}
\label{sec:conclusion}
The Gross-Pitaevskii-Poisson (GPP) equation encompasses the limits of both the nonlinear Schr\"{o}dinger equation and the Schr\"{o}dinger-Poisson equation. The former, in the absence of self-gravity, models superfluid $^4{\rm He}$ in the weak interaction regime and is widely employed to study superfluid turbulence~\cite{GK_2012_PhysRevE,Abid_2003,Kobayashi_2005_PhysRevLett}, characterized by a tangle of quantum vortices. 
Adding self-gravity to superfluid models significantly changes the dynamics of the system and extends their relevance across a wide range of length scales. The Gross-Pitaevskii-Poisson (GPP) model for self-gravitating superfluids has applications in phenomena within neutron stars ($\sim 10$ km), where it helps explain pulsar glitches~\cite{Shukla_2024_PhysRevD,Warszawski_2011}, and in modelling dark matter halos surrounding galaxies~\cite{Shukla_2024_PhysRevD.109.063009,chavanis_2011,Suarez}. Despite the broad applicability of the GPP equation, studies exploring its turbulence and spectra in self-gravitating superfluids remain limited. 

In our study, we perform direct numerical simulations of the GPP equation, without incorporating external forcing or dissipation, using a single control parameter, the ratio $l=\tfrac{\xi}{\lambda_{\rm J}}$, which determines the strength of the gravitational interaction. For varying values of $l$, the system undergoes a sequential collapse—initially forming sheet-like (pancake) structures, then transitioning into cylindrical shapes, and eventually evolving into a spherical configuration. We show that the appearance of these structures is a consequence of the anisotropic nature of gravitational instability. The gravitational collapse leads to the development of a minimum in the kinetic energy spectrum at the wavenumber $k_{\rm th}$. This minimum corresponds to the emergence of large-scale collapsed structures, introducing a characteristic length scale approximately equal diffusive scale of the objects. 

We show that for $k>k_{\rm th}$, the compressible kinetic energy spectrum follows $E_{\rm kin}^c \sim k^{2}$ suggesting the onset of thermalisation. In the uncollapsed regime, thermalisation occurs only at late times, around $t=90$ s in our DNS results. However, upon collapse, we observe an early onset of thermalisation, driven by the reduced relaxation time scale because of the long-range gravitational interaction. The number of particles, being the conserved quantity in the GPP Hamiltonian, decreases for $k>k_{\rm th}$ and moves towards the large scales. We associate this behavior of particle transfer from small to large scales with the inverse cascade because of the formation of condensates at large scales. This particle transfer mechanism aligns with the process of dark matter halo formation around galaxies, which the GPP model effectively captures. Using direct numerical simulations of the GPP equation, we reveal that the particle number spectrum follows a power-law distribution with an exponent of $-11/3$.

\section*{Acknowledgments}
SS thanks Rahul Pandit, Marc Brachet, and Kiran Kolluru for useful discussions and comments. We thank the Indo-French Centre for Applied Mathematics (IFCAM), Anusandhan National Research Foundation (ANRF), and the National Supercomputing Mission (NSM), India for support, and the Supercomputer Education and Research Centre (IISc) for computational resources.


\begin{thebibliography}{47}%
\makeatletter
\providecommand \@ifxundefined [1]{%
 \@ifx{#1\undefined}
}%
\providecommand \@ifnum [1]{%
 \ifnum #1\expandafter \@firstoftwo
 \else \expandafter \@secondoftwo
 \fi
}%
\providecommand \@ifx [1]{%
 \ifx #1\expandafter \@firstoftwo
 \else \expandafter \@secondoftwo
 \fi
}%
\providecommand \natexlab [1]{#1}%
\providecommand \enquote  [1]{``#1''}%
\providecommand \bibnamefont  [1]{#1}%
\providecommand \bibfnamefont [1]{#1}%
\providecommand \citenamefont [1]{#1}%
\providecommand \href@noop [0]{\@secondoftwo}%
\providecommand \href [0]{\begingroup \@sanitize@url \@href}%
\providecommand \@href[1]{\@@startlink{#1}\@@href}%
\providecommand \@@href[1]{\endgroup#1\@@endlink}%
\providecommand \@sanitize@url [0]{\catcode `\\12\catcode `\$12\catcode `\&12\catcode `\#12\catcode `\^12\catcode `\_12\catcode `\%12\relax}%
\providecommand \@@startlink[1]{}%
\providecommand \@@endlink[0]{}%
\providecommand \url  [0]{\begingroup\@sanitize@url \@url }%
\providecommand \@url [1]{\endgroup\@href {#1}{\urlprefix }}%
\providecommand \urlprefix  [0]{URL }%
\providecommand \Eprint [0]{\href }%
\providecommand \doibase [0]{https://doi.org/}%
\providecommand \selectlanguage [0]{\@gobble}%
\providecommand \bibinfo  [0]{\@secondoftwo}%
\providecommand \bibfield  [0]{\@secondoftwo}%
\providecommand \translation [1]{[#1]}%
\providecommand \BibitemOpen [0]{}%
\providecommand \bibitemStop [0]{}%
\providecommand \bibitemNoStop [0]{.\EOS\space}%
\providecommand \EOS [0]{\spacefactor3000\relax}%
\providecommand \BibitemShut  [1]{\csname bibitem#1\endcsname}%
\let\auto@bib@innerbib\@empty
\bibitem [{\citenamefont {Cichowlas}\ \emph {et~al.}(2005)\citenamefont {Cichowlas}, \citenamefont {Bona\"{\i}ti}, \citenamefont {Debbasch},\ and\ \citenamefont {Brachet}}]{Cyril_2005_PhysRevLett}%
  \BibitemOpen
  \bibfield  {author} {\bibinfo {author} {\bibfnamefont {C.}~\bibnamefont {Cichowlas}}, \bibinfo {author} {\bibfnamefont {P.}~\bibnamefont {Bona\"{\i}ti}}, \bibinfo {author} {\bibfnamefont {F.}~\bibnamefont {Debbasch}},\ and\ \bibinfo {author} {\bibfnamefont {M.}~\bibnamefont {Brachet}},\ }\bibfield  {title} {\bibinfo {title} {Effective dissipation and turbulence in spectrally truncated euler flows},\ }\href {https://doi.org/10.1103/PhysRevLett.95.264502} {\bibfield  {journal} {\bibinfo  {journal} {\href{https://link.aps.org/doi/10.1103/PhysRevLett.95.264502}{Phys. Rev. Lett.}}\ }\textbf {\bibinfo {volume} {95}},\ \bibinfo {pages} {264502} (\bibinfo {year} {2005})}\BibitemShut {NoStop}%
\bibitem [{\citenamefont {Maurer}\ and\ \citenamefont {Tabeling}(1998)}]{J_Maurer_1998}%
  \BibitemOpen
  \bibfield  {author} {\bibinfo {author} {\bibfnamefont {J.}~\bibnamefont {Maurer}}\ and\ \bibinfo {author} {\bibfnamefont {P.}~\bibnamefont {Tabeling}},\ }\bibfield  {title} {\bibinfo {title} {Local investigation of superfluid turbulence},\ }\href {https://doi.org/10.1209/epl/i1998-00314-9} {\bibfield  {journal} {\bibinfo  {journal} {Europhysics Letters}\ }\textbf {\bibinfo {volume} {43}},\ \bibinfo {pages} {29} (\bibinfo {year} {1998})}\BibitemShut {NoStop}%
\bibitem [{\citenamefont {Paoletti}\ and\ \citenamefont {Lathrop}(2011)}]{annurev_paoletti}%
  \BibitemOpen
  \bibfield  {author} {\bibinfo {author} {\bibfnamefont {M.~S.}\ \bibnamefont {Paoletti}}\ and\ \bibinfo {author} {\bibfnamefont {D.~P.}\ \bibnamefont {Lathrop}},\ }\bibfield  {title} {\bibinfo {title} {Quantum turbulence},\ }\href {https://doi.org/https://doi.org/10.1146/annurev-conmatphys-062910-140533} {\bibfield  {journal} {\bibinfo  {journal} {Annual Review of Condensed Matter Physics}\ }\textbf {\bibinfo {volume} {2}},\ \bibinfo {pages} {213} (\bibinfo {year} {2011})}\BibitemShut {NoStop}%
\bibitem [{\citenamefont {Henn}\ \emph {et~al.}(2009)\citenamefont {Henn}, \citenamefont {Seman}, \citenamefont {Roati}, \citenamefont {Magalh\~aes},\ and\ \citenamefont {Bagnato}}]{Henn_2009_PRL}%
  \BibitemOpen
  \bibfield  {author} {\bibinfo {author} {\bibfnamefont {E.~A.~L.}\ \bibnamefont {Henn}}, \bibinfo {author} {\bibfnamefont {J.~A.}\ \bibnamefont {Seman}}, \bibinfo {author} {\bibfnamefont {G.}~\bibnamefont {Roati}}, \bibinfo {author} {\bibfnamefont {K.~M.~F.}\ \bibnamefont {Magalh\~aes}},\ and\ \bibinfo {author} {\bibfnamefont {V.~S.}\ \bibnamefont {Bagnato}},\ }\bibfield  {title} {\bibinfo {title} {Emergence of turbulence in an oscillating bose-einstein condensate},\ }\href {https://doi.org/10.1103/PhysRevLett.103.045301} {\bibfield  {journal} {\bibinfo  {journal} {Phys. Rev. Lett.}\ }\textbf {\bibinfo {volume} {103}},\ \bibinfo {pages} {045301} (\bibinfo {year} {2009})}\BibitemShut {NoStop}%
\bibitem [{\citenamefont {Berloff}\ \emph {et~al.}(2014)\citenamefont {Berloff}, \citenamefont {Brachet},\ and\ \citenamefont {Proukakis}}]{Natalia_2014}%
  \BibitemOpen
  \bibfield  {author} {\bibinfo {author} {\bibfnamefont {N.~G.}\ \bibnamefont {Berloff}}, \bibinfo {author} {\bibfnamefont {M.}~\bibnamefont {Brachet}},\ and\ \bibinfo {author} {\bibfnamefont {N.~P.}\ \bibnamefont {Proukakis}},\ }\bibfield  {title} {\bibinfo {title} {Modeling quantum fluid dynamics at nonzero temperatures},\ }\href {https://doi.org/10.1073/pnas.1312549111} {\bibfield  {journal} {\bibinfo  {journal} {Proceedings of the National Academy of Sciences}\ }\textbf {\bibinfo {volume} {111}},\ \bibinfo {pages} {4675} (\bibinfo {year} {2014})}\BibitemShut {NoStop}%
\bibitem [{\citenamefont {Kobayashi}\ and\ \citenamefont {Tsubota}(2005)}]{Kobayashi_2005_PhysRevLett}%
  \BibitemOpen
  \bibfield  {author} {\bibinfo {author} {\bibfnamefont {M.}~\bibnamefont {Kobayashi}}\ and\ \bibinfo {author} {\bibfnamefont {M.}~\bibnamefont {Tsubota}},\ }\bibfield  {title} {\bibinfo {title} {Kolmogorov spectrum of superfluid turbulence: Numerical analysis of the gross-pitaevskii equation with a small-scale dissipation},\ }\href {https://doi.org/10.1103/PhysRevLett.94.065302} {\bibfield  {journal} {\bibinfo  {journal} {Phys. Rev. Lett.}\ }\textbf {\bibinfo {volume} {94}},\ \bibinfo {pages} {065302} (\bibinfo {year} {2005})}\BibitemShut {NoStop}%
\bibitem [{\citenamefont {Kobayashi}\ and\ \citenamefont {Tsubota}(2006)}]{Kobayashi_2006_PRL}%
  \BibitemOpen
  \bibfield  {author} {\bibinfo {author} {\bibfnamefont {M.}~\bibnamefont {Kobayashi}}\ and\ \bibinfo {author} {\bibfnamefont {M.}~\bibnamefont {Tsubota}},\ }\bibfield  {title} {\bibinfo {title} {Thermal dissipation in quantum turbulence},\ }\href {https://doi.org/10.1103/PhysRevLett.97.145301} {\bibfield  {journal} {\bibinfo  {journal} {Phys. Rev. Lett.}\ }\textbf {\bibinfo {volume} {97}},\ \bibinfo {pages} {145301} (\bibinfo {year} {2006})}\BibitemShut {NoStop}%
\bibitem [{\citenamefont {Krstulovic}(2012)}]{GK_2012_PhysRevE}%
  \BibitemOpen
  \bibfield  {author} {\bibinfo {author} {\bibfnamefont {G.}~\bibnamefont {Krstulovic}},\ }\bibfield  {title} {\bibinfo {title} {Kelvin-wave cascade and dissipation in low-temperature superfluid vortices},\ }\href {https://doi.org/10.1103/PhysRevE.86.055301} {\bibfield  {journal} {\bibinfo  {journal} {Phys. Rev. E}\ }\textbf {\bibinfo {volume} {86}},\ \bibinfo {pages} {055301} (\bibinfo {year} {2012})}\BibitemShut {NoStop}%
\bibitem [{\citenamefont {Verma}\ \emph {et~al.}(2022)\citenamefont {Verma}, \citenamefont {Pandit},\ and\ \citenamefont {Brachet}}]{AK_verma_2022}%
  \BibitemOpen
  \bibfield  {author} {\bibinfo {author} {\bibfnamefont {A.~K.}\ \bibnamefont {Verma}}, \bibinfo {author} {\bibfnamefont {R.}~\bibnamefont {Pandit}},\ and\ \bibinfo {author} {\bibfnamefont {M.~E.}\ \bibnamefont {Brachet}},\ }\bibfield  {title} {\bibinfo {title} {Rotating self-gravitating bose-einstein condensates with a crust: A model for pulsar glitches},\ }\href {https://doi.org/10.1103/PhysRevResearch.4.013026} {\bibfield  {journal} {\bibinfo  {journal} {Phys. Rev. Res.}\ }\textbf {\bibinfo {volume} {4}},\ \bibinfo {pages} {013026} (\bibinfo {year} {2022})}\BibitemShut {NoStop}%
\bibitem [{\citenamefont {Drummond}\ and\ \citenamefont {Melatos}(2017)}]{Drummond_2017_b}%
  \BibitemOpen
  \bibfield  {author} {\bibinfo {author} {\bibfnamefont {L.~V.}\ \bibnamefont {Drummond}}\ and\ \bibinfo {author} {\bibfnamefont {A.}~\bibnamefont {Melatos}},\ }\bibfield  {title} {\bibinfo {title} {{Stability of interlinked neutron vortex and proton flux-tube arrays in a neutron star – II. Far-from-equilibrium dynamics}},\ }\href {https://doi.org/10.1093/mnras/stx3197} {\bibfield  {journal} {\bibinfo  {journal} {Monthly Notices of the Royal Astronomical Society}\ }\textbf {\bibinfo {volume} {475}},\ \bibinfo {pages} {910} (\bibinfo {year} {2017})}\BibitemShut {NoStop}%
\bibitem [{\citenamefont {Shukla}\ \emph {et~al.}(2024{\natexlab{a}})\citenamefont {Shukla}, \citenamefont {Brachet},\ and\ \citenamefont {Pandit}}]{Shukla_2024_PhysRevD}%
  \BibitemOpen
  \bibfield  {author} {\bibinfo {author} {\bibfnamefont {S.}~\bibnamefont {Shukla}}, \bibinfo {author} {\bibfnamefont {M.~E.}\ \bibnamefont {Brachet}},\ and\ \bibinfo {author} {\bibfnamefont {R.}~\bibnamefont {Pandit}},\ }\bibfield  {title} {\bibinfo {title} {Neutron-superfluid vortices and proton-superconductor flux tubes: Development of a minimal model for pulsar glitches},\ }\href {https://doi.org/10.1103/PhysRevD.110.083002} {\bibfield  {journal} {\bibinfo  {journal} {Phys. Rev. D}\ }\textbf {\bibinfo {volume} {110}},\ \bibinfo {pages} {083002} (\bibinfo {year} {2024}{\natexlab{a}})}\BibitemShut {NoStop}%
\bibitem [{\citenamefont {Chavanis}(2011)}]{chavanis_2011}%
  \BibitemOpen
  \bibfield  {author} {\bibinfo {author} {\bibfnamefont {P.-H.}\ \bibnamefont {Chavanis}},\ }\href {https://doi.org/10.1103/PhysRevD.84.043531} {\bibfield  {journal} {\bibinfo  {journal} {Phys. Rev. D}\ }\textbf {\bibinfo {volume} {84}},\ \bibinfo {pages} {043531} (\bibinfo {year} {2011})}\BibitemShut {NoStop}%
\bibitem [{\citenamefont {Madarassy}\ and\ \citenamefont {Toth}(2015)}]{Madarassy_2015_PhysRevD}%
  \BibitemOpen
  \bibfield  {author} {\bibinfo {author} {\bibfnamefont {E.~J.~M.}\ \bibnamefont {Madarassy}}\ and\ \bibinfo {author} {\bibfnamefont {V.~T.}\ \bibnamefont {Toth}},\ }\bibfield  {title} {\bibinfo {title} {Evolution and dynamical properties of bose-einstein condensate dark matter stars},\ }\href {https://doi.org/10.1103/PhysRevD.91.044041} {\bibfield  {journal} {\bibinfo  {journal} {Phys. Rev. D}\ }\textbf {\bibinfo {volume} {91}},\ \bibinfo {pages} {044041} (\bibinfo {year} {2015})}\BibitemShut {NoStop}%
\bibitem [{\citenamefont {Su\'arez}\ \emph {et~al.}(2014)\citenamefont {Su\'arez}, \citenamefont {Robles},\ and\ \citenamefont {Matos}}]{Suarez}%
  \BibitemOpen
  \bibfield  {author} {\bibinfo {author} {\bibfnamefont {A.}~\bibnamefont {Su\'arez}}, \bibinfo {author} {\bibfnamefont {V.~H.}\ \bibnamefont {Robles}},\ and\ \bibinfo {author} {\bibfnamefont {T.}~\bibnamefont {Matos}},\ }\bibfield  {title} {\bibinfo {title} {{A Review on the Scalar Field/Bose-Einstein Condensate Dark Matter Model}},\ }\href {https://doi.org/10.1007/978-3-319-02063-1_9} {\bibfield  {journal} {\bibinfo  {journal} {Astrophys. Space Sci. Proc.}\ }\textbf {\bibinfo {volume} {38}},\ \bibinfo {pages} {107} (\bibinfo {year} {2014})},\ \Eprint {https://arxiv.org/abs/1302.0903} {arXiv:1302.0903 [astro-ph.CO]} \BibitemShut {NoStop}%
\bibitem [{\citenamefont {Chavanis}(2018)}]{chavanis_2018_PRD}%
  \BibitemOpen
  \bibfield  {author} {\bibinfo {author} {\bibfnamefont {P.-H.}\ \bibnamefont {Chavanis}},\ }\bibfield  {title} {\bibinfo {title} {Phase transitions between dilute and dense axion stars},\ }\href {https://doi.org/10.1103/PhysRevD.98.023009} {\bibfield  {journal} {\bibinfo  {journal} {Phys. Rev. D}\ }\textbf {\bibinfo {volume} {98}},\ \bibinfo {pages} {023009} (\bibinfo {year} {2018})}\BibitemShut {NoStop}%
\bibitem [{\citenamefont {Shukla}\ \emph {et~al.}(2024{\natexlab{b}})\citenamefont {Shukla}, \citenamefont {Verma}, \citenamefont {Brachet},\ and\ \citenamefont {Pandit}}]{Shukla_2024_PhysRevD.109.063009}%
  \BibitemOpen
  \bibfield  {author} {\bibinfo {author} {\bibfnamefont {S.}~\bibnamefont {Shukla}}, \bibinfo {author} {\bibfnamefont {A.~K.}\ \bibnamefont {Verma}}, \bibinfo {author} {\bibfnamefont {M.~E.}\ \bibnamefont {Brachet}},\ and\ \bibinfo {author} {\bibfnamefont {R.}~\bibnamefont {Pandit}},\ }\bibfield  {title} {\bibinfo {title} {Gravity- and temperature-driven phase transitions in a model for collapsed axionic condensates},\ }\href {https://doi.org/10.1103/PhysRevD.109.063009} {\bibfield  {journal} {\bibinfo  {journal} {Phys. Rev. D}\ }\textbf {\bibinfo {volume} {109}},\ \bibinfo {pages} {063009} (\bibinfo {year} {2024}{\natexlab{b}})}\BibitemShut {NoStop}%
\bibitem [{\citenamefont {Skipp}\ \emph {et~al.}(2020)\citenamefont {Skipp}, \citenamefont {L'vov},\ and\ \citenamefont {Nazarenko}}]{Jonathan_2020_PhysRevA}%
  \BibitemOpen
  \bibfield  {author} {\bibinfo {author} {\bibfnamefont {J.}~\bibnamefont {Skipp}}, \bibinfo {author} {\bibfnamefont {V.}~\bibnamefont {L'vov}},\ and\ \bibinfo {author} {\bibfnamefont {S.}~\bibnamefont {Nazarenko}},\ }\bibfield  {title} {\bibinfo {title} {Wave turbulence in self-gravitating bose gases and nonlocal nonlinear optics},\ }\href {https://doi.org/10.1103/PhysRevA.102.043318} {\bibfield  {journal} {\bibinfo  {journal} {Phys. Rev. A}\ }\textbf {\bibinfo {volume} {102}},\ \bibinfo {pages} {043318} (\bibinfo {year} {2020})}\BibitemShut {NoStop}%
\bibitem [{\citenamefont {Koplik}\ and\ \citenamefont {Levine}(1993)}]{Koplik_1993_PRL}%
  \BibitemOpen
  \bibfield  {author} {\bibinfo {author} {\bibfnamefont {J.}~\bibnamefont {Koplik}}\ and\ \bibinfo {author} {\bibfnamefont {H.}~\bibnamefont {Levine}},\ }\bibfield  {title} {\bibinfo {title} {Vortex reconnection in superfluid helium},\ }\href {https://doi.org/10.1103/PhysRevLett.71.1375} {\bibfield  {journal} {\bibinfo  {journal} {Phys. Rev. Lett.}\ }\textbf {\bibinfo {volume} {71}},\ \bibinfo {pages} {1375} (\bibinfo {year} {1993})}\BibitemShut {NoStop}%
\bibitem [{\citenamefont {Koplik}\ and\ \citenamefont {Levine}(1996)}]{Koplik_1996_PRL}%
  \BibitemOpen
  \bibfield  {author} {\bibinfo {author} {\bibfnamefont {J.}~\bibnamefont {Koplik}}\ and\ \bibinfo {author} {\bibfnamefont {H.}~\bibnamefont {Levine}},\ }\bibfield  {title} {\bibinfo {title} {Scattering of superfluid vortex rings},\ }\href {https://doi.org/10.1103/PhysRevLett.76.4745} {\bibfield  {journal} {\bibinfo  {journal} {Phys. Rev. Lett.}\ }\textbf {\bibinfo {volume} {76}},\ \bibinfo {pages} {4745} (\bibinfo {year} {1996})}\BibitemShut {NoStop}%
\bibitem [{\citenamefont {Leadbeater}\ \emph {et~al.}(2001)\citenamefont {Leadbeater}, \citenamefont {Winiecki}, \citenamefont {Samuels}, \citenamefont {Barenghi},\ and\ \citenamefont {Adams}}]{Leadbeater_2001_PRL}%
  \BibitemOpen
  \bibfield  {author} {\bibinfo {author} {\bibfnamefont {M.}~\bibnamefont {Leadbeater}}, \bibinfo {author} {\bibfnamefont {T.}~\bibnamefont {Winiecki}}, \bibinfo {author} {\bibfnamefont {D.~C.}\ \bibnamefont {Samuels}}, \bibinfo {author} {\bibfnamefont {C.~F.}\ \bibnamefont {Barenghi}},\ and\ \bibinfo {author} {\bibfnamefont {C.~S.}\ \bibnamefont {Adams}},\ }\bibfield  {title} {\bibinfo {title} {Sound emission due to superfluid vortex reconnections},\ }\href {https://doi.org/10.1103/PhysRevLett.86.1410} {\bibfield  {journal} {\bibinfo  {journal} {Phys. Rev. Lett.}\ }\textbf {\bibinfo {volume} {86}},\ \bibinfo {pages} {1410} (\bibinfo {year} {2001})}\BibitemShut {NoStop}%
\bibitem [{\citenamefont {Kerr}(2011)}]{Kerr_2011_PRL}%
  \BibitemOpen
  \bibfield  {author} {\bibinfo {author} {\bibfnamefont {R.~M.}\ \bibnamefont {Kerr}},\ }\bibfield  {title} {\bibinfo {title} {Vortex stretching as a mechanism for quantum kinetic energy decay},\ }\href {https://doi.org/10.1103/PhysRevLett.106.224501} {\bibfield  {journal} {\bibinfo  {journal} {Phys. Rev. Lett.}\ }\textbf {\bibinfo {volume} {106}},\ \bibinfo {pages} {224501} (\bibinfo {year} {2011})}\BibitemShut {NoStop}%
\bibitem [{\citenamefont {Rorai}\ \emph {et~al.}(2013)\citenamefont {Rorai}, \citenamefont {Sreenivasan},\ and\ \citenamefont {Fisher}}]{Rorai_2013}%
  \BibitemOpen
  \bibfield  {author} {\bibinfo {author} {\bibfnamefont {C.}~\bibnamefont {Rorai}}, \bibinfo {author} {\bibfnamefont {K.~R.}\ \bibnamefont {Sreenivasan}},\ and\ \bibinfo {author} {\bibfnamefont {M.~E.}\ \bibnamefont {Fisher}},\ }\bibfield  {title} {\bibinfo {title} {Propagating and annihilating vortex dipoles in the gross-pitaevskii equation},\ }\href {https://doi.org/10.1103/PhysRevB.88.134522} {\bibfield  {journal} {\bibinfo  {journal} {Phys. Rev. B}\ }\textbf {\bibinfo {volume} {88}},\ \bibinfo {pages} {134522} (\bibinfo {year} {2013})}\BibitemShut {NoStop}%
\bibitem [{\citenamefont {Kobayashi}\ \emph {et~al.}(2021)\citenamefont {Kobayashi}, \citenamefont {Parnaudeau}, \citenamefont {Luddens}, \citenamefont {Lothodé}, \citenamefont {Danaila}, \citenamefont {Brachet},\ and\ \citenamefont {Danaila}}]{KOBAYASHI2021107579}%
  \BibitemOpen
  \bibfield  {author} {\bibinfo {author} {\bibfnamefont {M.}~\bibnamefont {Kobayashi}}, \bibinfo {author} {\bibfnamefont {P.}~\bibnamefont {Parnaudeau}}, \bibinfo {author} {\bibfnamefont {F.}~\bibnamefont {Luddens}}, \bibinfo {author} {\bibfnamefont {C.}~\bibnamefont {Lothodé}}, \bibinfo {author} {\bibfnamefont {L.}~\bibnamefont {Danaila}}, \bibinfo {author} {\bibfnamefont {M.}~\bibnamefont {Brachet}},\ and\ \bibinfo {author} {\bibfnamefont {I.}~\bibnamefont {Danaila}},\ }\bibfield  {title} {\bibinfo {title} {Quantum turbulence simulations using the gross–pitaevskii equation: High-performance computing and new numerical benchmarks},\ }\href {https://doi.org/https://doi.org/10.1016/j.cpc.2020.107579} {\bibfield  {journal} {\bibinfo  {journal} {Computer Physics Communications}\ }\textbf {\bibinfo {volume} {258}},\ \bibinfo {pages} {107579} (\bibinfo {year} {2021})}\BibitemShut {NoStop}%
\bibitem [{\citenamefont {Shukla}\ \emph {et~al.}(2025)\citenamefont {Shukla}, \citenamefont {Krstulovic},\ and\ \citenamefont {Pandit}}]{shukla2024capturereleasequantumvortices}%
  \BibitemOpen
  \bibfield  {author} {\bibinfo {author} {\bibfnamefont {S.}~\bibnamefont {Shukla}}, \bibinfo {author} {\bibfnamefont {G.}~\bibnamefont {Krstulovic}},\ and\ \bibinfo {author} {\bibfnamefont {R.}~\bibnamefont {Pandit}},\ }\bibfield  {title} {\bibinfo {title} {Capture and release of quantum vortices using mechanical devices in low-temperature superfluids},\ }\href {https://doi.org/10.1103/PhysRevB.111.L100504} {\bibfield  {journal} {\bibinfo  {journal} {Phys. Rev. B}\ }\textbf {\bibinfo {volume} {111}},\ \bibinfo {pages} {L100504} (\bibinfo {year} {2025})}\BibitemShut {NoStop}%
\bibitem [{\citenamefont {Feynman}(1955)}]{FEYNMAN195517}%
  \BibitemOpen
  \bibfield  {author} {\bibinfo {author} {\bibfnamefont {R.}~\bibnamefont {Feynman}},\ }\bibfield  {title} {\bibinfo {title} {Chapter ii application of quantum mechanics to liquid helium}\ }(\bibinfo  {publisher} {Elsevier},\ \bibinfo {year} {1955})\ pp.\ \bibinfo {pages} {17--53}\BibitemShut {NoStop}%
\bibitem [{\citenamefont {Ogawa}\ \emph {et~al.}(2002)\citenamefont {Ogawa}, \citenamefont {Tsubota},\ and\ \citenamefont {Hattori}}]{Ogawa_2002_JPS}%
  \BibitemOpen
  \bibfield  {author} {\bibinfo {author} {\bibfnamefont {S.-i.}\ \bibnamefont {Ogawa}}, \bibinfo {author} {\bibfnamefont {M.}~\bibnamefont {Tsubota}},\ and\ \bibinfo {author} {\bibfnamefont {Y.}~\bibnamefont {Hattori}},\ }\bibfield  {title} {\bibinfo {title} {Study of reconnection and acoustic emission of quantized vortices in superfluid by the numerical analysis of the gross–pitaevskii equation},\ }\href {https://doi.org/10.1143/JPSJ.71.813} {\bibfield  {journal} {\bibinfo  {journal} {Journal of the Physical Society of Japan}\ }\textbf {\bibinfo {volume} {71}},\ \bibinfo {pages} {813} (\bibinfo {year} {2002})}\BibitemShut {NoStop}%
\bibitem [{\citenamefont {Nore}\ \emph {et~al.}(1997{\natexlab{a}})\citenamefont {Nore}, \citenamefont {Abid},\ and\ \citenamefont {Brachet}}]{Nore_Marc_1997_POF}%
  \BibitemOpen
  \bibfield  {author} {\bibinfo {author} {\bibfnamefont {C.}~\bibnamefont {Nore}}, \bibinfo {author} {\bibfnamefont {M.}~\bibnamefont {Abid}},\ and\ \bibinfo {author} {\bibfnamefont {M.~E.}\ \bibnamefont {Brachet}},\ }\bibfield  {title} {\bibinfo {title} {Decaying kolmogorov turbulence in a model of superflow},\ }\href {https://doi.org/10.1063/1.869473} {\bibfield  {journal} {\bibinfo  {journal} {Physics of Fluids}\ }\textbf {\bibinfo {volume} {9}},\ \bibinfo {pages} {2644} (\bibinfo {year} {1997}{\natexlab{a}})},\ \Eprint {https://arxiv.org/abs/https://pubs.aip.org/aip/pof/article-pdf/9/9/2644/19175152/2644\_1\_online.pdf} {https://pubs.aip.org/aip/pof/article-pdf/9/9/2644/19175152/2644\_1\_online.pdf} \BibitemShut {NoStop}%
\bibitem [{\citenamefont {Nore}\ \emph {et~al.}(1997{\natexlab{b}})\citenamefont {Nore}, \citenamefont {Abid},\ and\ \citenamefont {Brachet}}]{Nore_1997_PhysRevLett}%
  \BibitemOpen
  \bibfield  {author} {\bibinfo {author} {\bibfnamefont {C.}~\bibnamefont {Nore}}, \bibinfo {author} {\bibfnamefont {M.}~\bibnamefont {Abid}},\ and\ \bibinfo {author} {\bibfnamefont {M.~E.}\ \bibnamefont {Brachet}},\ }\bibfield  {title} {\bibinfo {title} {Kolmogorov turbulence in low-temperature superflows},\ }\href {https://doi.org/10.1103/PhysRevLett.78.3896} {\bibfield  {journal} {\bibinfo  {journal} {\href{https://link.aps.org/doi/10.1103/PhysRevLett.78.3896}{Phys. Rev. Lett.}}\ }\textbf {\bibinfo {volume} {78}},\ \bibinfo {pages} {3896} (\bibinfo {year} {1997}{\natexlab{b}})}\BibitemShut {NoStop}%
\bibitem [{\citenamefont {Araki}\ \emph {et~al.}(2002)\citenamefont {Araki}, \citenamefont {Tsubota},\ and\ \citenamefont {Nemirovskii}}]{Araki_2002_PRL}%
  \BibitemOpen
  \bibfield  {author} {\bibinfo {author} {\bibfnamefont {T.}~\bibnamefont {Araki}}, \bibinfo {author} {\bibfnamefont {M.}~\bibnamefont {Tsubota}},\ and\ \bibinfo {author} {\bibfnamefont {S.~K.}\ \bibnamefont {Nemirovskii}},\ }\bibfield  {title} {\bibinfo {title} {Energy spectrum of superfluid turbulence with no normal-fluid component},\ }\href {https://doi.org/10.1103/PhysRevLett.89.145301} {\bibfield  {journal} {\bibinfo  {journal} {Phys. Rev. Lett.}\ }\textbf {\bibinfo {volume} {89}},\ \bibinfo {pages} {145301} (\bibinfo {year} {2002})}\BibitemShut {NoStop}%
\bibitem [{\citenamefont {{Polanco}}\ \emph {et~al.}(2021)\citenamefont {{Polanco}}, \citenamefont {{M{\"u}ller}},\ and\ \citenamefont {{Krstulovic}}}]{Polanco_2021NatComm}%
  \BibitemOpen
  \bibfield  {author} {\bibinfo {author} {\bibfnamefont {J.~I.}\ \bibnamefont {{Polanco}}}, \bibinfo {author} {\bibfnamefont {N.~P.}\ \bibnamefont {{M{\"u}ller}}},\ and\ \bibinfo {author} {\bibfnamefont {G.}~\bibnamefont {{Krstulovic}}},\ }\bibfield  {title} {\bibinfo {title} {{Vortex clustering, polarisation and circulation intermittency in classical and quantum turbulence}},\ }\href {https://doi.org/10.1038/s41467-021-27382-6} {\bibfield  {journal} {\bibinfo  {journal} {Nature Communications}\ }\textbf {\bibinfo {volume} {12}},\ \bibinfo {eid} {7090} (\bibinfo {year} {2021})},\ \Eprint {https://arxiv.org/abs/2107.03335} {arXiv:2107.03335 [physics.flu-dyn]} \BibitemShut {NoStop}%
\bibitem [{\citenamefont {Frisch}(1995)}]{Frisch_1995}%
  \BibitemOpen
  \bibfield  {author} {\bibinfo {author} {\bibfnamefont {U.}~\bibnamefont {Frisch}},\ }\href@noop {} {\emph {\bibinfo {title} {Turbulence: The Legacy of A. N. Kolmogorov}}}\ (\bibinfo  {publisher} {Cambridge University Press},\ \bibinfo {year} {1995})\BibitemShut {NoStop}%
\bibitem [{\citenamefont {Chan}\ \emph {et~al.}(2012)\citenamefont {Chan}, \citenamefont {Mitra},\ and\ \citenamefont {Brandenburg}}]{Axel_2012_PhysRevE}%
  \BibitemOpen
  \bibfield  {author} {\bibinfo {author} {\bibfnamefont {C.-k.}\ \bibnamefont {Chan}}, \bibinfo {author} {\bibfnamefont {D.}~\bibnamefont {Mitra}},\ and\ \bibinfo {author} {\bibfnamefont {A.}~\bibnamefont {Brandenburg}},\ }\bibfield  {title} {\bibinfo {title} {Dynamics of saturated energy condensation in two-dimensional turbulence},\ }\href {https://doi.org/10.1103/PhysRevE.85.036315} {\bibfield  {journal} {\bibinfo  {journal} {Phys. Rev. E}\ }\textbf {\bibinfo {volume} {85}},\ \bibinfo {pages} {036315} (\bibinfo {year} {2012})}\BibitemShut {NoStop}%
\bibitem [{\citenamefont {Boffetta}\ and\ \citenamefont {Ecke}(2012)}]{Boffetta_2012}%
  \BibitemOpen
  \bibfield  {author} {\bibinfo {author} {\bibfnamefont {G.}~\bibnamefont {Boffetta}}\ and\ \bibinfo {author} {\bibfnamefont {R.~E.}\ \bibnamefont {Ecke}},\ }\bibfield  {title} {\bibinfo {title} {Two-dimensional turbulence},\ }\href {https://doi.org/https://doi.org/10.1146/annurev-fluid-120710-101240} {\bibfield  {journal} {\bibinfo  {journal} {Annual Review of Fluid Mechanics}\ }\textbf {\bibinfo {volume} {44}},\ \bibinfo {pages} {427} (\bibinfo {year} {2012})}\BibitemShut {NoStop}%
\bibitem [{\citenamefont {Kiessling}(2003)}]{KIESSLING_2003}%
  \BibitemOpen
  \bibfield  {author} {\bibinfo {author} {\bibfnamefont {M.~K.-H.}\ \bibnamefont {Kiessling}},\ }\bibfield  {title} {\bibinfo {title} {The “jeans swindle”: A true story—mathematically speaking},\ }\href {https://doi.org/https://doi.org/10.1016/S0196-8858(02)00556-0} {\bibfield  {journal} {\bibinfo  {journal} {Advances in Applied Mathematics}\ }\textbf {\bibinfo {volume} {31}},\ \bibinfo {pages} {132} (\bibinfo {year} {2003})}\BibitemShut {NoStop}%
\bibitem [{\citenamefont {Falco}\ \emph {et~al.}(2013)\citenamefont {Falco}, \citenamefont {Hansen}, \citenamefont {Wojtak},\ and\ \citenamefont {Mamon}}]{Falco_2013}%
  \BibitemOpen
  \bibfield  {author} {\bibinfo {author} {\bibfnamefont {M.}~\bibnamefont {Falco}}, \bibinfo {author} {\bibfnamefont {S.~H.}\ \bibnamefont {Hansen}}, \bibinfo {author} {\bibfnamefont {R.}~\bibnamefont {Wojtak}},\ and\ \bibinfo {author} {\bibfnamefont {G.~A.}\ \bibnamefont {Mamon}},\ }\bibfield  {title} {\bibinfo {title} {{Why does the Jeans Swindle work?}},\ }\href {https://doi.org/10.1093/mnrasl/sls051} {\bibfield  {journal} {\bibinfo  {journal} {Monthly Notices of the Royal Astronomical Society: Letters}\ }\textbf {\bibinfo {volume} {431}},\ \bibinfo {pages} {L6} (\bibinfo {year} {2013})}\BibitemShut {NoStop}%
\bibitem [{\citenamefont {Proukakis}\ and\ \citenamefont {Jackson}(2008)}]{Proukakis_2008}%
  \BibitemOpen
  \bibfield  {author} {\bibinfo {author} {\bibfnamefont {N.~P.}\ \bibnamefont {Proukakis}}\ and\ \bibinfo {author} {\bibfnamefont {B.}~\bibnamefont {Jackson}},\ }\bibfield  {title} {\bibinfo {title} {Finite-temperature models of bose–einstein condensation},\ }\href@noop {} {\bibfield  {journal} {\bibinfo  {journal} {\href{https://dx.doi.org/10.1088/0953-4075/41/20/203002}{Journal of Physics B: Atomic, Molecular and Optical Physics}}\ }\textbf {\bibinfo {volume} {41}},\ \bibinfo {pages} {203002} (\bibinfo {year} {2008})}\BibitemShut {NoStop}%
\bibitem [{\citenamefont {Sin}(1994)}]{Sin_PhysRevD_1994}%
  \BibitemOpen
  \bibfield  {author} {\bibinfo {author} {\bibfnamefont {S.-J.}\ \bibnamefont {Sin}},\ }\bibfield  {title} {\bibinfo {title} {Late-time phase transition and the galactic halo as a bose liquid},\ }\href@noop {} {\bibfield  {journal} {\bibinfo  {journal} {\href{https://link.aps.org/doi/10.1103/PhysRevD.50.3650}{Phys. Rev. D}}\ }\textbf {\bibinfo {volume} {50}},\ \bibinfo {pages} {3650} (\bibinfo {year} {1994})}\BibitemShut {NoStop}%
\bibitem [{\citenamefont {{Sofue}}\ and\ \citenamefont {{Rubin}}(2001)}]{Rubin_araa}%
  \BibitemOpen
  \bibfield  {author} {\bibinfo {author} {\bibfnamefont {Y.}~\bibnamefont {{Sofue}}}\ and\ \bibinfo {author} {\bibfnamefont {V.}~\bibnamefont {{Rubin}}},\ }\bibfield  {title} {\bibinfo {title} {{Rotation Curves of Spiral Galaxies}},\ }\href {https://doi.org/10.1146/annurev.astro.39.1.137} {\bibfield  {journal} {\bibinfo  {journal} {araa}\ }\textbf {\bibinfo {volume} {39}},\ \bibinfo {pages} {137} (\bibinfo {year} {2001})},\ \Eprint {https://arxiv.org/abs/astro-ph/0010594} {arXiv:astro-ph/0010594 [astro-ph]} \BibitemShut {NoStop}%
\bibitem [{\citenamefont {Hou}\ and\ \citenamefont {Li}(2007)}]{HOU_dealiasing}%
  \BibitemOpen
  \bibfield  {author} {\bibinfo {author} {\bibfnamefont {T.~Y.}\ \bibnamefont {Hou}}\ and\ \bibinfo {author} {\bibfnamefont {R.}~\bibnamefont {Li}},\ }\href {https://doi.org/https://doi.org/10.1016/j.jcp.2007.04.014} {\bibfield  {journal} {\bibinfo  {journal} {Journal of Computational Physics}\ }\textbf {\bibinfo {volume} {226}},\ \bibinfo {pages} {379} (\bibinfo {year} {2007})}\BibitemShut {NoStop}%
\bibitem [{\citenamefont {Taylor}\ and\ \citenamefont {Green}(1937)}]{Taylor_1937}%
  \BibitemOpen
  \bibfield  {author} {\bibinfo {author} {\bibfnamefont {G.~I.}\ \bibnamefont {Taylor}}\ and\ \bibinfo {author} {\bibfnamefont {A.~E.}\ \bibnamefont {Green}},\ }\bibfield  {title} {\bibinfo {title} {Mechanism of the production of small eddies from large ones},\ }\href {https://doi.org/10.1098/rspa.1937.0036} {\bibfield  {journal} {\bibinfo  {journal} {Proceedings of the Royal Society of London. Series A - Mathematical and Physical Sciences}\ }\textbf {\bibinfo {volume} {158}},\ \bibinfo {pages} {499} (\bibinfo {year} {1937})}\BibitemShut {NoStop}%
\bibitem [{\citenamefont {{Zel'dovich}}(1970)}]{Zeldovich}%
  \BibitemOpen
  \bibfield  {author} {\bibinfo {author} {\bibfnamefont {Y.~B.}\ \bibnamefont {{Zel'dovich}}},\ }\bibfield  {title} {\bibinfo {title} {{Gravitational instability: An approximate theory for large density perturbations.}},\ }\href@noop {} {\bibfield  {journal} {\bibinfo  {journal} {AAP}\ }\textbf {\bibinfo {volume} {5}},\ \bibinfo {pages} {84} (\bibinfo {year} {1970})}\BibitemShut {NoStop}%
\bibitem [{\citenamefont {{Peebles}}(1980)}]{PeeblesP}%
  \BibitemOpen
  \bibfield  {author} {\bibinfo {author} {\bibfnamefont {P.~J.~E.}\ \bibnamefont {{Peebles}}},\ }\href@noop {} {\emph {\bibinfo {title} {{The large-scale structure of the universe}}}}\ (\bibinfo {year} {1980})\BibitemShut {NoStop}%
\bibitem [{\citenamefont {Barenghi}\ \emph {et~al.}(2014)\citenamefont {Barenghi}, \citenamefont {Skrbek},\ and\ \citenamefont {Sreenivasan}}]{Barenghi_2014_pnas}%
  \BibitemOpen
  \bibfield  {author} {\bibinfo {author} {\bibfnamefont {C.~F.}\ \bibnamefont {Barenghi}}, \bibinfo {author} {\bibfnamefont {L.}~\bibnamefont {Skrbek}},\ and\ \bibinfo {author} {\bibfnamefont {K.~R.}\ \bibnamefont {Sreenivasan}},\ }\bibfield  {title} {\bibinfo {title} {Introduction to quantum turbulence},\ }\href {https://doi.org/10.1073/pnas.1400033111} {\bibfield  {journal} {\bibinfo  {journal} {Proceedings of the National Academy of Sciences}\ }\textbf {\bibinfo {volume} {111}},\ \bibinfo {pages} {4647} (\bibinfo {year} {2014})},\ \Eprint {https://arxiv.org/abs/https://www.pnas.org/doi/pdf/10.1073/pnas.1400033111} {https://www.pnas.org/doi/pdf/10.1073/pnas.1400033111} \BibitemShut {NoStop}%
\bibitem [{\citenamefont {Krstulovic}\ and\ \citenamefont {Brachet}(2011)}]{Giorgio_thermalisation}%
  \BibitemOpen
  \bibfield  {author} {\bibinfo {author} {\bibfnamefont {G.}~\bibnamefont {Krstulovic}}\ and\ \bibinfo {author} {\bibfnamefont {M.}~\bibnamefont {Brachet}},\ }\bibfield  {title} {\bibinfo {title} {Dispersive bottleneck delaying thermalization of turbulent bose-einstein condensates},\ }\href {https://doi.org/10.1103/PhysRevLett.106.115303} {\bibfield  {journal} {\bibinfo  {journal} {Phys. Rev. Lett.}\ }\textbf {\bibinfo {volume} {106}},\ \bibinfo {pages} {115303} (\bibinfo {year} {2011})}\BibitemShut {NoStop}%
\bibitem [{\citenamefont {Kirkpatrick}\ \emph {et~al.}(2020)\citenamefont {Kirkpatrick}, \citenamefont {Mirasola},\ and\ \citenamefont {Prescod-Weinstein}}]{Kay_PRD_2020}%
  \BibitemOpen
  \bibfield  {author} {\bibinfo {author} {\bibfnamefont {K.}~\bibnamefont {Kirkpatrick}}, \bibinfo {author} {\bibfnamefont {A.~E.}\ \bibnamefont {Mirasola}},\ and\ \bibinfo {author} {\bibfnamefont {C.}~\bibnamefont {Prescod-Weinstein}},\ }\bibfield  {title} {\bibinfo {title} {Relaxation times for bose-einstein condensation in axion miniclusters},\ }\href {https://doi.org/10.1103/PhysRevD.102.103012} {\bibfield  {journal} {\bibinfo  {journal} {Phys. Rev. D}\ }\textbf {\bibinfo {volume} {102}},\ \bibinfo {pages} {103012} (\bibinfo {year} {2020})}\BibitemShut {NoStop}%
\bibitem [{\citenamefont {Abid}\ \emph {et~al.}(2003)\citenamefont {Abid}, \citenamefont {Huepe}, \citenamefont {Metens}, \citenamefont {Nore}, \citenamefont {Pham}, \citenamefont {Tuckerman},\ and\ \citenamefont {Brachet}}]{Abid_2003}%
  \BibitemOpen
  \bibfield  {author} {\bibinfo {author} {\bibfnamefont {M.}~\bibnamefont {Abid}}, \bibinfo {author} {\bibfnamefont {C.}~\bibnamefont {Huepe}}, \bibinfo {author} {\bibfnamefont {S.}~\bibnamefont {Metens}}, \bibinfo {author} {\bibfnamefont {C.}~\bibnamefont {Nore}}, \bibinfo {author} {\bibfnamefont {C.~T.}\ \bibnamefont {Pham}}, \bibinfo {author} {\bibfnamefont {L.~S.}\ \bibnamefont {Tuckerman}},\ and\ \bibinfo {author} {\bibfnamefont {M.~E.}\ \bibnamefont {Brachet}},\ }\bibfield  {title} {\bibinfo {title} {Gross–pitaevskii dynamics of bose–einstein condensates and superfluid turbulence},\ }\href {https://doi.org/10.1016/j.fluiddyn.2003.09.001} {\bibfield  {journal} {\bibinfo  {journal} {Fluid Dynamics Research}\ }\textbf {\bibinfo {volume} {33}},\ \bibinfo {pages} {509} (\bibinfo {year} {2003})}\BibitemShut {NoStop}%
\bibitem [{\citenamefont {Warszawski}\ and\ \citenamefont {Melatos}(2011)}]{Warszawski_2011}%
  \BibitemOpen
  \bibfield  {author} {\bibinfo {author} {\bibfnamefont {L.}~\bibnamefont {Warszawski}}\ and\ \bibinfo {author} {\bibfnamefont {A.}~\bibnamefont {Melatos}},\ }\bibfield  {title} {\bibinfo {title} {{Gross–Pitaevskii model of pulsar glitches}},\ }\href {https://doi.org/10.1111/j.1365-2966.2011.18803.x} {\bibfield  {journal} {\bibinfo  {journal} {Monthly Notices of the Royal Astronomical Society}\ }\textbf {\bibinfo {volume} {415}},\ \bibinfo {pages} {1611} (\bibinfo {year} {2011})}\BibitemShut {NoStop}%
\end{thebibliography}

%

\end{document}